\documentclass[aps, prb, 10pt, twocolumn, superscriptaddress, notitlepage, floatfix, longbibliography]{revtex4-2}
\usepackage[T1]{fontenc}
\usepackage[utf8]{inputenc}
\usepackage{bm}
\usepackage{microtype}
\usepackage{indentfirst}
\usepackage{graphicx}
\usepackage{adjustbox}

\usepackage{amsmath, amssymb, amsthm}
\usepackage{mathtools}
\DeclarePairedDelimiter{\abs}{\lvert}{\rvert}
\DeclarePairedDelimiter{\kh}{\lparen}{\rparen}
\DeclarePairedDelimiter{\fkh}{\lbrack}{\rbrack}
\DeclarePairedDelimiter{\hkh}{\lbrace}{\rbrace}
\makeatletter
\let\oldabs\abs
\def\abs{\@ifstar{\oldabs}{\oldabs*}}
\let\oldkh\kh
\def\kh{\@ifstar{\oldkh}{\oldkh*}}
\let\oldfkh\fkh
\def\fkh{\@ifstar{\oldfkh}{\oldfkh*}}
\let\oldhkh\hkh
\def\hkh{\@ifstar{\oldhkh}{\oldhkh*}}
\makeatother

\DeclareMathOperator{\Tr}{Tr}

\DeclareMathOperator{\Rep}{Rep}

\usepackage{dsfont}

\def\C{\mathds{C}}
\def\Z{\mathds{Z}}

\usepackage{braket}
\newcommand{\op}[1]{\ket{#1}\!\bra{#1}}
\newcommand{\dket}[1]{\ket{#1}\!\rangle}
\newcommand{\dbra}[1]{\langle\!\bra{#1}}
\newcommand{\dbraket}[1]{\langle\!\braket{#1}\!\rangle}
\newcommand{\ten}{\gamma_{\textrm{EN}}}
\newcommand{\tmi}{\gamma_{\textrm{MI}}}

\newcommand{\euler}{\,\mathrm{e}}
\newcommand{\ii}{\mathrm{i}}

\newcommand{\vb}[1]{\bm{#1}}

\newcommand{\cA}{\mathcal{A}}
\newcommand{\cD}{\mathcal{D}}
\newcommand{\cH}{\mathcal{H}}
\newcommand{\cO}{\mathcal{O}}

\usepackage{tikz}
\usetikzlibrary{arrows.meta}
\usetikzlibrary{decorations.markings}
\tikzset{arrowmid/.style={decoration={markings, mark= at position 0.5 with {\arrow{stealth}}}, postaction={decorate}}}

\usepackage{xcolor}

\usepackage{hyperref}

\begin{document}
\title{Entanglement negativity in decohered topological states}
\author{Kang-Le Cai}
\author{Meng Cheng}
\affiliation{Department of Physics, Yale University, New Haven, Connecticut 06511, USA}

\begin{abstract}
  We investigate universal entanglement signatures of mixed-state phases obtained by decohering pure-state topological order (TO), focusing on topological corrections to logarithmic entanglement negativity and mutual information: topological entanglement negativity (TEN) and topological mutual information (TMI). For Abelian TOs under decoherence, we develop a replica field-theory framework based on a doubled-state construction that relates TEN and TMI to the quantum dimensions of domain-wall defects between decoherence-induced topological boundary conditions, yielding general expressions in the strong-decoherence regime. We further compute TEN and TMI exactly for decohered $G$-graded string-net states, including cases with non-Abelian anyons. We interpret the results within the strong one-form-symmetry framework for mixed-state TOs: TMI probes the total quantum dimension of the emergent premodular anyon theory, whereas TEN detects only its modular part.
\end{abstract}

\maketitle

\section{Introduction}

Recently, there has been increasing interest in mixed-state phases of matter, particularly those obtained by decohering pure-state topological orders (TOs)~\cite{Dennis:2001nw, Coser2019classificationof, Bao:2023zry, Fan:2023rvp, Ellison:2024svg, wang2025intrinsic, Sohal:2024qvq, Li:2024rgz, Luo_unpub, Ogata:2025otz}. This interest is driven by the fact that realistic experimental platforms are inevitably noisy. Consequently, a key question is: what universal quantities can be extracted from these mixed states to characterize them?

It is insightful to draw inspiration from pure-state TOs, which emerge in ground states of gapped local Hamiltonians. A celebrated result due to Kitaev--Preskill~\cite{Kitaev:2005dm} and Levin--Wen~\cite{levin2006detecting} is that the entanglement entropy $S_A$ of a large connected region $A$ exhibits a topological correction:
\begin{equation}\label{Eq_Intro_TEE}
  S_A = \alpha \abs{\partial A} - \gamma + \cdots,
\end{equation}
where $\gamma$ is the topological entanglement entropy (TEE), and $\dots$ denotes contributions that vanish as the linear size of $A$ becomes large. It is believed that in generic models, $\gamma$ is given by $\ln \cD$, where $\cD$ is the total quantum dimension, a topological invariant of the phase. An important caveat is that in some models the value of $\gamma$ can be greater than $\ln \cD$, a phenomenon known as spurious TEE~\cite{Zou:2016dck, Williamson:2018zig, KatoPRR2020}. It has recently been proven that $\ln \cD$ is the universal lower bound on $\gamma$~\cite{Kim:2023ydi, Levin:2024ngk}.

It is natural to seek similar universal corrections in entanglement measures for mixed-state TOs. Generally speaking, characterizing (bipartite) entanglement in mixed states is more challenging than in pure states. The entanglement entropy and mutual information receive contributions from both quantum entanglement and classical correlations and therefore are no longer entanglement monotones. Correspondingly, the so-called topological mutual information (TMI), constructed from mutual information as a mixed-state analogue of TEE, may contain both quantum and classical contributions.

By contrast, the entanglement negativity (EN)~\cite{Vidal:2002zz} more cleanly captures quantum entanglement in mixed states: it vanishes for bipartite separable states and is an entanglement monotone~\cite{Plenio:2005cwa}. Generalizing Eq.~\eqref{Eq_Intro_TEE} to many-body mixed states, one can define topological entanglement negativity (TEN), with the hope that TEN captures the essential contribution of quantum entanglement in mixed-state phases.

Recently, Ref.~\cite{Fan:2023rvp} studied TEN in a $\Z_2$ toric code under bit-flip noise and showed that TEN can indeed distinguish the mixed-state TO from the trivial state. When the error rate $p$ is below the decoding threshold $p_c$, TEN is equal to $\ln 2$, the same value as in the pure-state toric code. For $p>p_c$, TEN vanishes. Thus in this case, TEN serves as a good indicator of the topological phase and the phase transition.

In this work, we systematically compute TEN and TMI in two general classes of decohered topological states. First, we consider Abelian topological phases subject to decoherence channels that proliferate anyons. We employ a field-theoretical formulation, in which the computation of TEN and TMI is mapped to finding the quantum dimensions of certain topological defects on the boundary of a three-dimensional topological quantum field theory. This allows us to derive general expressions for TEN and TMI in this case.

The next family of examples we study is decohered $G$-graded string-net states, where $G$ is a finite group. By $G$ grading we mean that the pure-state TO has a $G$ gauge structure. In particular, it has a set of anyons labeled by irreducible representations of $G$, whose collection is denoted by $\Rep(G)$. We then analytically compute TEN and TMI in the mixed state obtained by proliferating $\Rep(G)$ anyons. Importantly, in this case both the parent TO and the proliferated anyons can be non-Abelian.

We then interpret the results from the perspective of generalized one-form symmetries of the mixed-state TO~\cite{Ellison:2024svg, Sohal:2024qvq, wang2025intrinsic}. Based on our calculations, we discuss possible universal lower bounds for TEN in terms of the modular one-form symmetries.

The paper is organized as follows. In Sec.~\ref{Sec_EntanglementMeasure}, we review the relevant entanglement measures. In Sec.~\ref{Sec_TQFT}, we present our field-theoretic calculation of TEN and TMI for decohered Abelian TOs. Sec.~\ref{Sec_GradedStringNet} focuses on TEN and TMI in the decohered $G$-graded string-net models. Finally, in Sec.~\ref{Sec_1form}, we discuss the one-form symmetry characterization of mixed-state TOs and relate TEN and TMI to quantum dimensions of the anyon theory of the mixed-state TO.

\section{Entanglement measures in mixed states}\label{Sec_EntanglementMeasure}

We primarily focus on two types of entanglement measures: the von Neumann entanglement entropy and the logarithmic entanglement negativity, along with their R\'enyi versions, which are employed for computational convenience. In this section, we briefly review the definitions of these entanglement measures and their relevance in characterizing topological phases.

For a quantum state $\rho$, the von Neumann entanglement entropy of a region $A$ is defined as
\begin{equation}
  S_A(\rho) = - \Tr \rho_A \ln\rho_A,
\end{equation}
where the reduced density matrix $\rho_A$ is obtained by tracing out the complement $\bar{A}$, i.e., $\rho_A = \Tr_{\bar{A}} \rho$. Using the replica trick, the entanglement entropy can be expressed as the analytic continuation of the R\'enyi entanglement entropy:
\begin{equation}
  S_A(\rho) = \lim_{n\to 1} S_A^{(n)}(\rho),
\end{equation}
where $S_A^{(n)}(\rho)$ denotes the R\'enyi entanglement entropy of order $n$, defined by
\begin{equation}
  S_A^{(n)}(\rho) = \frac{1}{1-n} \ln \Tr \rho_A^n.
\end{equation}

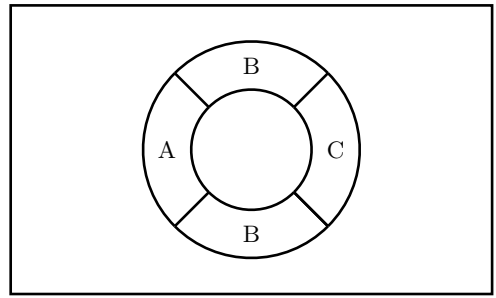
\begin{figure}[ht]
  \centering
  \begin{tikzpicture}[scale=0.8, line width=1pt]
    \def\rin{1cm}
    \def\rout{1.8cm}
    \foreach \startangle/\endangle in { -45/45, 45/135, 135/225, 225/315 } {
      \draw (\startangle:\rout) arc (\startangle:\endangle:\rout);
      \draw (\endangle:\rin) arc (\endangle:\startangle:\rin);
    \draw (\startangle:\rout) -- (\startangle:\rin);}
    \draw (315:\rout) -- (315:\rin);
    \node at (90:1.4cm) {B};
    \node at (270:1.4cm) {B};
    \node at (180:1.4cm) {A};
    \node at (0:1.4cm) {C};
    \draw (-4,-2.4) rectangle (4,2.4);
  \end{tikzpicture}
  \caption{Levin--Wen geometry for extracting the topological correction in $S_A$. The regions $A$, $B$, and $C$ are chosen such that the volume-law and area-law contributions cancel out in Eq.~\eqref{Eq_EM_CMI}.}
  \label{Fig_TMIAnnulus}
\end{figure}

For gapped ground states in two dimensions, the entanglement entropy of a large connected region is expected to exhibit leading area-law scaling with a subleading constant correction. In a generic setup, this constant correction can contain nonuniversal contributions, such as those arising from corners in the geometry of the region. The Levin--Wen construction, illustrated in Fig.~\ref{Fig_TMIAnnulus}, is designed to isolate the topological part by forming a linear combination of entanglement entropies in which the area-law term and other local contributions cancel~\cite{levin2006detecting}. The resulting TEE $\gamma$ is extracted as
\begin{equation}\label{Eq_EM_CMI}
  \gamma = \frac12 ( S_{AB} + S_{BC} - S_B - S_{ABC} ),
\end{equation}
in the limit of large $A,B,C$.

For decohered gapped states, the entanglement entropy typically exhibits leading volume-law scaling. Nevertheless, the Levin--Wen combination in Eq.~\eqref{Eq_EM_CMI} cancels the volume-law term as well, and the remainder defines the TMI, denoted $\tmi$.

Originally introduced in Ref.~\cite{Vidal:2002zz}, logarithmic negativity is an entanglement measure based on the Peres–-Horodecki criterion. It provides an efficiently computable upper bound on the distillable entanglement.

For a quantum state $\rho$, the logarithmic negativity associated with region $A$ is defined as
\begin{equation}
  \mathcal{E}_A(\rho) = \ln \| \rho^{\intercal_A} \|,
\end{equation}
where $\rho^{\intercal_A}$ denotes the partial transpose of $\rho$ with respect to region $A$, and $\| \cdot \|$ denotes the trace norm.

As an entanglement measure, logarithmic negativity possesses several desirable properties~\cite{Vidal:2002zz, Plenio:2005cwa}:
\begin{enumerate}
  \item $\mathcal{E}_A(\rho)$ vanishes when $\rho$ is separable (although the converse is not true).
  \item $\mathcal{E}_A(\rho)$ is an entanglement monotone; that is, it does not increase under local operations and classical communication (LOCC).
  \item For pure states $\rho = \op{\psi}$, the logarithmic negativity $\mathcal{E}_A(\rho)$ reduces to the R\'enyi entropy $S_A^{(\frac12)}(\rho)$. In this sense, it serves as a generalization of entanglement entropy to mixed states.
\end{enumerate}

In many cases (e.g., in a quantum field theory), computing the logarithmic negativity directly is not feasible. Analogous to the entanglement entropy, the logarithmic negativity can be obtained using the replica trick, which expresses it as the analytic continuation of the R\'enyi entanglement negativity:
\begin{equation}
  \mathcal{E}_A(\rho) = \lim_{\textrm{even }n \to 1} \mathcal{E}_A^{(n)}(\rho),
\end{equation}
where $\mathcal{E}_A^{(n)}(\rho)$ denotes the R\'enyi entanglement negativity of order $n$, defined for even integers $n$ as
\begin{equation}
  \mathcal{E}_A^{(n)}(\rho) = \frac{1}{2-n} \ln \frac{ \Tr (\rho^{\intercal_A})^n }{ \Tr \rho^n }.
\end{equation}

In contrast to entanglement entropy, it is generally expected that the leading scaling of entanglement negativity obeys an area law even in decohered topological states. TEN $\ten$ is defined as the constant correction to the entanglement negativity $\mathcal{E}_A(\rho)$ in the limit of large $A$.

We note that there exist other entanglement measures for mixed states that are more informative than logarithmic negativity. However, they are typically much more difficult to compute, as they often involve nontrivial minimization procedures. A notable example is the entanglement of formation (EF), which is studied for mixed-state TOs in Ref.~\cite{Lessa:2025xut}.

\section{Field-theoretic calculation for decohered Abelian TO}\label{Sec_TQFT}

In this section, we develop a field-theoretic framework for computing TEN and TMI of decohered Abelian TOs in the strong decoherence limit.

\subsection{Topological entanglement negativity}\label{Sec_TQFT_TEN}

We begin by introducing the doubled-state construction, which provides a compact formulation of the R\'enyi negativity. Given a density matrix $\rho$, we define its corresponding doubled state $\dket{\rho}$ in the doubled Hilbert space $\cH \otimes \cH^*$ as
\begin{equation}
  \rho = \sum_{ij} \rho_{ij} \ket{i}\bra{j} \Longrightarrow
  \dket{\rho} = \sum_{ij} \rho_{ij} \ket{i} \ket{j^*},
\end{equation}
where $\hkh{\ket{i}}$ is an orthonormal basis for the Hilbert space $\cH$, and we use $*$ to keep track of states in the bra space $\cH^*$. When a bipartition $A \cup \bar{A}$ is specified, we choose $\hkh{\ket{i}}$ to be compatible with the tensor product structure $\cH = \cH_A \otimes \cH_{\bar{A}}$, i.e., $\ket{i} = \ket{i_A} \otimes \ket{i_{\bar{A}}}$.

For the $n$-fold tensor product $\rho^{\otimes n} = \rho \otimes \cdots \otimes \rho$, we define its doubled state $\dket{\rho^{\otimes n}}$ in the $n$-replica space $(\cH \otimes \cH^*)^{\otimes n}$ as
\begin{equation}
  \dket{\rho^{\otimes n}} = \sum_{i,j} \rho_{i_1,j_1} \cdots \rho_{i_n,j_n} \ket{i_1} \ket{j_1^*} \cdots \ket{i_n} \ket{j_n^*}.
\end{equation}

We also introduce a reference state $\dket{\mathcal{E}_n}$, defined by
\begin{equation}
  \dket{\mathcal{E}_n} = \sum_i \ket{i_1} \ket{i_2^*} \ket{i_2} \ket{i_3^*} \cdots \ket{i_n} \ket{i_1^*}.
\end{equation}
Then, one can show that
\begin{equation}
  \begin{split}\label{Eq_TQFT_Doubled}
    \Tr \rho^n  &= \dbraket{ \mathcal{E}_n | \rho^{\otimes n} }, \\
    \Tr (\rho^{\intercal_A})^n &= \dbraket{ \mathcal{E}_n | (R_A^*)^2 | \rho^{\otimes n}},
  \end{split}
\end{equation}
where $R_A^*$ cyclically permutes region $A$ in the $\cH^*$ Hilbert space of the replicas. Explicitly, the action of $(R_A^*)^2$ on the reference state yields
\begin{equation}
  \begin{split}
    (R_A^*)^2 \dket{\mathcal{E}_n} = \sum_i & \ket{i_1} \ket{ i_{n,A}^* , i_{2,\bar{A}}^* } \ket{i_2} \ket{ i_{1,A}^* , i_{3,\bar{A}}^* } \\
    & \quad \cdots \ket{i_n} \ket{ i_{n-1,A}^* , i_{1,\bar{A}}^* }.
  \end{split}
\end{equation}
The R\'enyi negativity of $\rho$ is then given by
\begin{equation}
  \mathcal{E}_A^{(n)}(\rho) = \frac{1}{2-n} \ln \frac{ \dbraket{ \mathcal{E}_n | (R_A^*)^2 | \rho^{\otimes n} } }{ \dbraket{ \mathcal{E}_n | \rho^{\otimes n} } }.
\end{equation}

\begin{figure}[ht]
  \centering
  \includegraphics[width=0.6\linewidth]{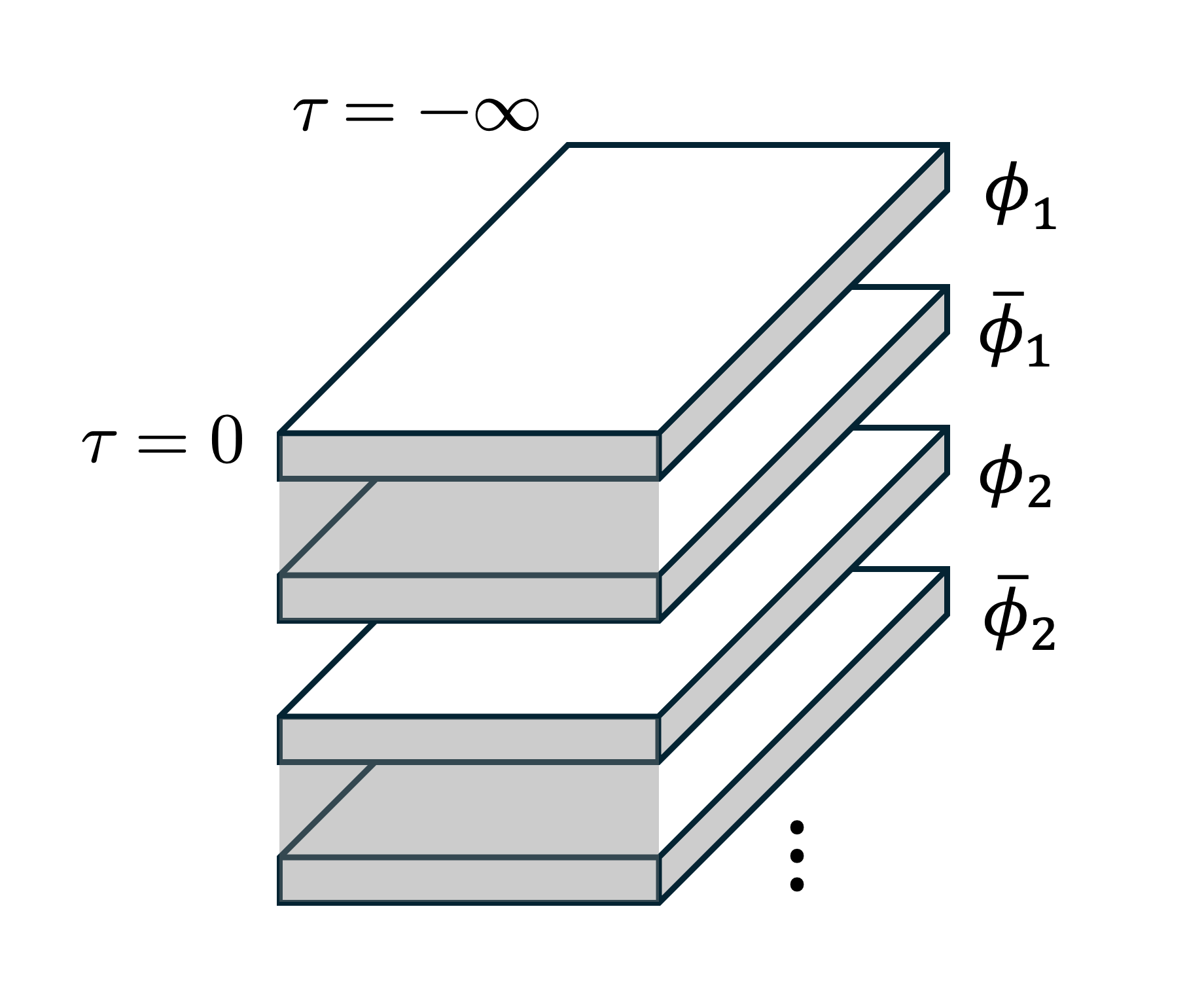}
  \caption{Path-integral representation of Eq.~\eqref{Eq_TQFT_Doubled}. Without decoherence, the reference bra states $\dbra{\mathcal{E}_n}$ and $\dbra{\mathcal{E}_n} (R_A^*)^2$ impose corresponding boundary conditions at the $\tau = 0$ surface. The effect of decoherence can be captured by interactions (represented by the grey region) that couple $\phi_i$ and $\bar{\phi}_i$ at $\tau=0$.}
  \label{Fig_PathIntegral}
\end{figure}

We assume that the pure-state TO under consideration is described by a topological quantum field theory (TQFT). In particular, Abelian TOs can always be described by Abelian Chern--Simons theories. In the absence of decoherence, we take $\rho$ to be the ground state of the TO. The replica ground state $\dket{\rho^{\otimes n}}$ can be prepared by a Euclidean path integral on $2n$ replicas from imaginary time $\tau=-\infty$ to $\tau=0$ (see Fig.~\ref{Fig_PathIntegral}).

The reference bra states $\dbra{\mathcal{E}_n}$ and $\dbra{\mathcal{E}_n} (R_A^*)^2$ in Eq.~\eqref{Eq_TQFT_Doubled} can be interpreted as specific boundary conditions for the path integral at $\tau=0$. To understand their effects more explicitly, suppose that the boundary theory at $\tau=0$ has fields $\phi_i$ and $\bar{\phi}_i$ (for the $i$-th replica). The reference bra state $\dbra{\mathcal{E}_n}$ then imposes the boundary condition $\bar{\phi}_i(\tau=0) = \phi_{i+1}(\tau=0)$. Concretely, the boundary theory can be a multi-component Luttinger liquid where $\phi$ and $\bar{\phi}$ are compact bosons.

The effect of decoherence (quantum channels) on $\dket{\rho^{\otimes n}}$ can be captured by boundary interactions at $\tau=0$ in the path integral, which drive the system to new boundary conditions. In particular, the decoherence-induced interactions couple only the $\phi_i$ and $\bar{\phi}_i$ modes.

Since the theory is topological, the bulk (of the Euclidean spacetime) can be uniquely associated with a modular tensor category (MTC), i.e., an anyon theory, denoted by $\mathcal{C}$. The replica state $\dket{\rho^{\otimes n}}$ corresponds to $(\mathcal{C} \boxtimes \mathcal{C}^*)^{\boxtimes n}$, where $\mathcal{C}^*$ is the MTC with all anyon data complex conjugated.

A key assumption we make is that the boundary state induced by decoherence is topological. Topological boundary conditions of an MTC are classified by the anyons condensed on the boundary~\cite{LevinPRX2013, Kong:2013aya}. Mathematically, these condensed anyons correspond to a Lagrangian algebra in the MTC. For Abelian TOs, the Lagrangian algebra admits a much simpler description in terms of Lagrangian subgroups. A Lagrangian subgroup is a set of bosonic anyons, closed under fusion, with trivial self and mutual statistics. The total number of such bosons must be equal to $\cD$, which is the total quantum dimension of the Abelian TO.

Consider the simplest case with $n=1$, corresponding to two layers $\mathcal{C} \boxtimes \mathcal{C}^*$. The anyons are denoted by $(a, b^*)$ for $a, b \in \mathcal{C}$, where the superscript $*$ indicates that the anyon $b^*$ resides in the $\mathcal{C}^*$ layer. There is a canonical boundary condition labeled by the following Lagrangian subgroup:
\begin{equation}
  \sum_{a \in \mathcal{C}}(a, \bar{a}^*),
\end{equation}
where $\bar{a}$ is the dual of $a$ (i.e., $a \times \bar{a} = 0$). This is the boundary condition set by the reference bra state $\dbra{\mathcal{E}_1}$.

Generalizing to $n$ replicas, $(\mathcal{C} \boxtimes \mathcal{C}^*)^{\boxtimes n}$, the reference state $\dbra{{\mathcal{E}}_n}$ leads to the following Lagrangian subgroup:
\begin{equation}
  \cA_0 = \sum_{a_1, \cdots, a_n \in \mathcal{C}} (a_1, \bar{a}_2^*, {a}_2, \bar{a}_3^*, \cdots, a_n, \bar{a}_1^*).
\end{equation}

Now, we turn on the decoherence-induced interactions. Here we make two further assumptions:
\begin{enumerate}
  \item We assume that the decoherence channel proliferates Abelian anyons that belong to a group $\mathcal{B} \subset \mathcal{C}$.
  \item We assume that the decoherence strength is strong; in particular, we assume that decoherence drives the boundary to a new topological boundary condition.
\end{enumerate}
Our next task is to determine the new topological boundary state. We have seen that there are two competing anyon condensations, one from the reference state and the other from decoherence. First, we can write down the subgroup of the decoherence-proliferated anyons as
\begin{equation}
  \begin{split}
    \cA_{\mathcal{B}} = & \sum_{a \in \mathcal{B}} (a, \bar{a}^*, 0, 0^*, \cdots, 0, 0^*) \\
    & + \sum_{a \in \mathcal{B}} (0, 0^*, a, \bar{a}^*, \cdots, 0, 0^*) + \cdots.
  \end{split}
\end{equation}
Throughout, for two subgroups $\cA$ and $\cA'$, we write $\cA + \cA'$ for the subgroup generated by $\cA$ and $\cA'$.

In the strong-decoherence limit, all condensed anyons in $\cA_0$ that braid nontrivially with $\cA_{\mathcal{B}}$ are suppressed. We denote by $\cA_0/\cA_{\cal B}$ the subgroup of anyons in $\cA_0$ which braid trivially with $\cA_{\cal B}$.  The new subgroup describing the boundary condition induced by decoherence is
\begin{equation}\label{Eq_TQFT_DecoheredBoundary}
  \cA_0' = \cA_0 / \cA_{\mathcal{B}} + \cA_{\mathcal{B}}.
\end{equation}

We can repeat this process for $(R_A^*)^2 \dket{{\mathcal{E}_n}}$, which corresponds to a different boundary algebra in region $A$:
\begin{equation}
  \cA_1 = (R^*)^2 \cA_0 = \sum_{a_1, \cdots, a_n \in \mathcal{C}} (a_1, \bar{a}_n^*, {a}_2, \bar{a}_1^*, \cdots, a_{n}, \bar{a}_{n-1}^*).
\end{equation}
Under the effect of decoherence, the new boundary condition is described by $\cA_1' = \cA_1 / \cA_{\mathcal{B}} + \cA_{\mathcal{B}}$.

\begin{figure}[ht]
  \centering

  \begin{tikzpicture}[line width=1pt, scale=0.8]
    \coordinate (A) at (0, 0, 0);
    \coordinate (B) at (2, 0, 0);
    \coordinate (C) at (2, 2, 0);
    \coordinate (D) at (0, 2, 0);
    \coordinate (E) at (0, 0, 5);
    \coordinate (F) at (2, 0, 5);
    \coordinate (G) at (2, 2, 5);
    \coordinate (H) at (0, 2, 5);

    \draw (B) -- (C) -- (D);
    \draw[dashed, line width=0.5pt, opacity=0.5] (D) -- (A) -- (B);
    \draw (E) -- (F) -- (G) -- (H) -- cycle;
    \draw[dashed, line width=0.5pt, opacity=0.5] (A) -- (E);
    \draw (B) -- (F);
    \draw (C) -- (G);
    \draw (D) -- (H);

    \draw[blue] (0,1,5) -- (2,1,5) -- (2,1,0);
    \draw[blue, dashed, opacity=0.5] (0,1,5) -- (0,1,0) -- (2,1,0);


    \draw (1,1,5) circle (0.5) node[above] {$A$};
    \fill[red] (0.5,1,5) circle (2pt);
    \fill[red] (1.5,1,5) circle (2pt);
  \end{tikzpicture}

  \caption{Schematic picture in which the $2n$ replicas are compressed into a single piece. The red dots denote the domain walls between two boundary conditions on a spatial slice.}
  \label{Fig_DomainWall}
\end{figure}
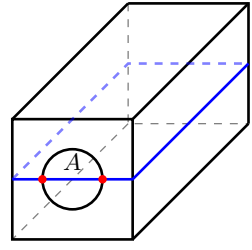

\begin{figure}[ht]
  \centering

  \begin{tikzpicture}[line width=1pt, scale=0.8]
    \draw (0,0) -- (9,0);

    \draw[rounded corners=0.3cm] (1,0) -- (1,2) -- (5,2) -- (5,1.12);
    \draw (5,0.88) -- (5,0);
    \draw[rounded corners=0.3cm] (3,0) -- (3,1) -- (7,1) -- (7,0);

    \fill[red] (2,0) circle (3pt);
    \fill[red] (4,0) circle (3pt);
    \fill[red] (6,0) circle (3pt);
    \fill[red] (8,0) circle (3pt);

    \node[below] at (1,0) {$\mathcal{A}_0'$};
    \node[below] at (3,0) {$\mathcal{A}_1'$};
    \node[below] at (5,0) {$\mathcal{A}_0'$};
    \node[below] at (7,0) {$\mathcal{A}_1'$};

    \node[above] at (3,2) {$W$};
    \node[above] at (6,1) {$\tilde{W}$};
  \end{tikzpicture}

  \caption{Configuration with four junctions between $\cA_0'$ and $\cA_1'$.}
  \label{Fig_StringOperator}
\end{figure}
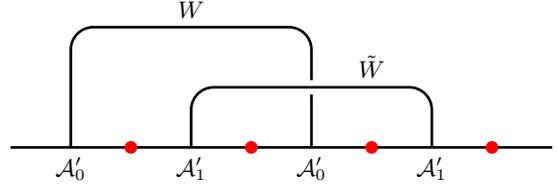

The negativity is associated with $(R^*_A)^2$, which can be interpreted as a topological defect operator that creates a domain wall between the $\mathcal{A}_0'$ and $\mathcal{A}_1'$ boundary conditions. In the TQFT limit, the expectation value of this defect operator is determined by the quantum dimension of the defect on the boundary.

To compute this defect quantum dimension, we take a slice of the theory perpendicular to the boundary and canonically quantize the theory on the slice (see Fig.~\ref{Fig_DomainWall} for an illustration). The edge of this slice is gapped with Lagrangian subgroup given by $\cA_0'$. Introducing $(R_A^*)^2$ then creates domain walls between $\cA_0'$ and $\cA_1'$.
The quantum dimension can be extracted from the topological ground-state degeneracy associated with these domain walls.

More concretely, the domain wall between $\cA_0'$ and $\cA_1'$ has a quantum dimension $d_n$, defined as follows: when there are $2m$ junctions arranged on a circle, the dimension of the ground-state space scales as $d_n^{2m}$. The R\'enyi negativity in the topological limit is then given by
\begin{equation}
  \mathcal{E}_A^{(n)} = \frac{1}{2-n} \ln d_n = -\ten.
\end{equation}

To compute the quantum dimension $d_n$, we study four junctions between $\cA_0'$ and $\cA_1'$, as shown in Fig.~\ref{Fig_StringOperator}. The boundary is divided into four segments $\cA_0', \cA_1', \cA_0'$ and $\cA_1'$. We denote the string operators between the two $\cA_0'$ segments by $W_b$, where $b$ is an anyon in $\cA_0'$. Similarly, the string operators between the two $\cA_1'$ segments are denoted by $W'_c$, where $c$ is an anyon in $\cA_1'$. The string operators $W_b$ and $W'_c$ satisfy nontrivial commutation relations determined by the braiding phase between the anyons $b$ and $c$.

We now make an additional assumption: the ground-state degeneracy associated with the domain walls is equal to the minimal dimension required by the algebra of string operators. That is, the degeneracy is purely topological. We cannot rule out the possibility of additional degeneracy protected by symmetries of the system (for example, there is an anti-unitary symmetry arising from the hermiticity of the density matrix), but we will ignore such possibilities in what follows. If such additional degeneracy were to arise, then our results only provide a lower bound to $\ten$.

The topologically degenerate ground states form an irreducible representation of this string operator algebra with the smallest possible dimension. In the Abelian case, it suffices to consider the commutation relations among the generators of the string operator algebra, which correspond to generators of the Lagrangian subgroup. In Appendix~\ref{Sec_RepTh}, we analyze the representation theory of such algebras, relating their representations to those of string operator algebras on a torus. We also review a concrete method for determining the dimension of the irreducible representation.

Let the minimal-dimension representation of the four-junction algebra have dimension $M_n$. For $2m$ junctions, there are $m-1$ independent copies of the four-junction algebra, leading to a minimal total dimension $M_n^{m-1}$. Consequently, the quantum dimension of the junction is given by $d_n = \sqrt{M_n}$, and we obtain
\begin{equation}
  \ten = \frac{1}{2(n-2)} \ln M_n.
\end{equation}

\subsection{Topological mutual information}

The same doubled-state construction can also be used to extract TMI. The R\'enyi entanglement entropy $S_A^{(n)}(\rho)$ can be computed by expressing $\Tr \rho_A^n$ as
\begin{equation}
  \Tr \rho_A^n = \dbraket{ \mathcal{E}_n | R_{\bar{A}}^* | \rho^{\otimes n} }.
\end{equation}

Within the field-theoretic framework, we extract $\tmi$ directly as the topological contribution to the entanglement entropy, rather than using the annulus geometry described in Eq.~\eqref{Eq_EM_CMI}. The relevant domain wall is that between $\cA_0' = \cA_0 / \cA_{\mathcal{B}} + \cA_{\mathcal{B}}$ and $\cA_2' = \cA_2 / \cA_{\mathcal{B}} + \cA_{\mathcal{B}}$, where
\begin{equation}
  \cA_2 = R^* \cA_0 = \sum_{a_1, \cdots, a_n \in \mathcal{C}} (a_1, \bar{a}_1^*, {a}_2, \bar{a}_2^*, \cdots, a_{n}, \bar{a}_n^*).
\end{equation}
A notable distinction is that all anyons in $\cA_2$ braid trivially with those in $\cA_{\mathcal{B}}$. Consequently, decoherence does not affect $\cA_2$, and we always have $\cA_2' = \cA_2$.

For a configuration with four junctions between $\cA_0'$ and $\cA_2$, let $W$ denote the string operators connecting the two $\cA_0'$ segments, and $W''$ those connecting the two $\cA_2$ segments. If the minimal-dimension representation of the string operator algebra generated by $W$ and $W''$ has dimension $M_n$, then
\begin{equation}
  \tmi = \frac{1}{2(n-1)} \ln M_n.
\end{equation}

\subsection{Pure state TO}\label{Sec_TQFT_PureTO}

In this section, we verify that our framework reproduces the expected TEN and TMI for a pure Abelian TO~\cite{LeeVidalTEN, CastelnovoTEN}:
\begin{equation}
  \ten = \tmi = \ln \cD,
\end{equation}
where $\cD$ is the total quantum dimension of the Abelian TO.

In the absence of decoherence, we have $\cA_0' = \cA_0$ and $\cA_1' = \cA_1$. Let $\mathcal{G}$ denote a generating set of anyons in the Abelian TO. The generators of the Lagrangian subgroup $\cA_0$ are labeled by $(j,b)$, where $j = 1, 2, \cdots, n$ and $b \in \mathcal{G}$. Specifically, $(j,b)$ corresponds to the anyon
\begin{equation}
  (a_1, \bar{a}_2^*, {a}_2, \bar{a}_3^*, \cdots, a_n, \bar{a}_1^*),
\end{equation}
where $a_j=b$ and $a_i=0$ for $i \neq j$. We denote the corresponding string operators by $W(j,b)$. In a similar manner, we construct string operators for $\cA_1$ and $\cA_2$, denoted $W'(j,b)$ and $W''(j,b)$, respectively.

To compute TEN, we focus on the nontrivial commutation relations between $W(j,b)$ and $W'(j,b)$:
\begin{equation}
  \begin{split}\label{Eq_TQFT_TEN_Alg}
    W(j,b) W'(j,c) & = B(b,c) W'(j,c) W(j,b), \\
    W(j+2,b) W'(j,c) & = B(b,c)^{-1} W'(j,c) W(j+2,b),
  \end{split}
\end{equation}
where $B(b,c)$ is the braiding phase between anyons $b, c \in \mathcal{G}$.

The algebra splits into two equivalent parts, corresponding to even and odd values of $j$, so it suffices to analyze one of them. For $j=1, 2, \cdots, n$, define
\begin{equation}
  T(j,b) =
  \begin{cases}
    W(j,b), & j\textrm{ odd}\\
    W'(j-1,b), & j\textrm{ even}
  \end{cases}.
\end{equation}
The operators $T(j,b)$ then satisfy
\begin{equation}\label{Eq_TQFT_Alg_Reduced}
  T(j,b) T(j+1,c) = B(b,c) T(j+1,c) T(j,b).
\end{equation}
The minimal-dimension representation of this algebra is closely related to the minimal-dimension representation of the string operator algebra of the Abelian TO on a torus, whose dimension is $\cD^2$. In Appendix~\ref{Sec_RepTh}, we show that the minimal-dimension representation of $T(j,b)$ has dimension $\cD^{n-2}$. Thus, we obtain $M_n = \cD^{2(n-2)}$ and $\ten = \ln \cD$, as expected.

Similarly, to compute TMI, we consider the nontrivial commutation relations between $W(j,b)$ and $W''(j,b)$:
\begin{equation}
  \begin{split}\label{Eq_TQFT_TMI_Alg}
    W(j,b) W''(j,c) & = B(b,c) W''(j,c) W(j,b),\\
    W(j+1,b) W''(j,c) & = B(b,c)^{-1} W''(j,c) W(j+1,b),
  \end{split}
\end{equation}
where $b, c \in \mathcal{G}$. We define
\begin{equation}
  T(j,b) =
  \begin{cases}
    W(\frac{j+1}{2},b), & j\textrm{ odd}\\
    W''(\frac{j}{2},b), & j\textrm{ even}
  \end{cases},
\end{equation}
with $j=1, 2, \cdots, 2n$. The commutation relations between these $T(j,b)$ are also given by Eq.~\eqref{Eq_TQFT_Alg_Reduced}, and the minimal-dimension representation now has dimension $M_n = \cD^{2(n-1)}$, which reproduces $\tmi = \ln \cD$.

\subsection{Decohered TO}\label{Sec_TQFT_Decohered}

We now consider TEN and TMI for a decohered Abelian TO obtained by proliferating a subgroup of anyons generated by a single anyon $x \in \mathcal{C}$. The anyons that braid trivially with $x$ form a subgroup, denoted by $\mathcal{C}_x$, which describes the anyon theory of the decohered state. We denote a generating set of $\mathcal{C}_x$ by $\mathcal{G}_x$.

To compute TEN, we enumerate a generating set of the Lagrangian subgroup $\cA_0' = \cA_0 / \cA_{\mathcal{B}} + \cA_{\mathcal{B}}$.

First, $\cA_{\mathcal{B}}$ has $n$ generators of the form
\begin{equation}
  (a_1, \bar{a}_1^*, {a}_2, \bar{a}_2^*, \cdots, a_{n}, \bar{a}_n^*),
\end{equation}
where exactly one $a_j=x$ and $a_i=0$ for $i \neq j$.

The other generators of $\cA_0'$ are from $\cA_0/\cA_{\mathcal{B}}$. Consider a general anyon in $\cA_0$ of the form
\begin{equation}\label{Eq_TEN_Anyon}
  (a_1, \bar{a}_2^*, {a}_2, \bar{a}_3^*, \cdots, a_{n}, \bar{a}_1^*),
\end{equation}
with $a_j \in \mathcal{C}$. For this anyon to braid trivially with the generators of $\cA_{\mathcal{B}}$, it must satisfy
\begin{equation}
  B(x,a_j) = B(x,a_{j+1}) \quad \forall j.
\end{equation}
Thus, $\cA_0 / \cA_{\mathcal{B}}$ decomposes into subsets, each labeled by the common value of $B(x,a_j)$. The subset with $B(x,a_j)=1$ forms a subgroup, whose generators can be chosen such that exactly one $a_j = b \in \mathcal{G}_x$ is nontrivial. We denote these generators by $(j,b)$ and the corresponding string operators by $W(j,b)$.

The remaining generators of $\cA_0 / \cA_{\mathcal{B}}$ can be chosen such that $a_j \equiv b$ for some $b$ that braids nontrivially with $x$. To see this, consider an anyon in $\cA_0 / \cA_{\mathcal{B}}$ with $B(x,a_j) \equiv \euler^{\ii\theta} \neq 1$. For each $j$, we have $B(x, a_1 \bar{a}_j)=1$. This allows us to set $a_j=a_1$ by fusing with the generators $(j,b)$. As a result, the detailed structure of these generators is not important for the calculation, since they braid trivially with all anyons in $\cA_0$, $\cA_1$, and $\cA_2$.

The generating set of $\cA_1'$ can be analyzed in an analogous way. We again have the $n$ generators of $\cA_{\mathcal{B}}$. For $\cA_1 / \cA_{\mathcal{B}}$, consider a general anyon in $\cA_1$ of the form
\begin{equation}
  (a_1, \bar{a}_n^*, a_2, \bar{a}_3^*, \cdots, a_{n}, \bar{a}_{n-1}^*),
\end{equation}
with $a_j \in \mathcal{C}$. For this anyon to braid trivially with the generators of $\cA_{\mathcal{B}}$, we must have
\begin{equation}
  B(x,a_j) = B(x,a_{j-1}) \quad \forall j.
\end{equation}
Again, the subset with $B(x,a_j)=1$ is a subgroup generated by anyons with exactly one nontrivial $a_j = c \in \mathcal{G}_x$. We denote the corresponding string operators by $W'(j,c)$. The remaining generators of $\cA_1 / \cA_{\mathcal{B}}$ can be chosen such that $a_j \equiv c$, so they coincide with those of $\cA_0 / \cA_{\mathcal{B}}$.

The string operators $W(j,b)$ and $W'(j,c)$ again satisfy the commutation relations in Eq.~\eqref{Eq_TQFT_TEN_Alg}, now with $b, c \in \mathcal{G}_x$. By the same logic as in the pure-state case, we obtain
\begin{equation}
  \ten = \ln \sqrt{\abs{\mathcal{M}_x}},
\end{equation}
where $\abs{\mathcal{M}_x}$ is the number of anyons in the modular part of $\mathcal{C}_x$. Transparent anyons do not contribute as they braid trivially with all other anyons by definition. Further support for this statement is provided in Appendix~\ref{Sec_RepTh_TwistedApplication}.

The case of TMI differs because $\cA_2$ remains unaffected by decoherence. The relevant string operators are still $W''(j,c)$, defined in Sec.~\ref{Sec_TQFT_PureTO}. The nontrivial commutation relations between $W(j,b)$ and $W''(j,c)$ are given by Eq.~\eqref{Eq_TQFT_TMI_Alg}, now with $b \in \mathcal{G}_x$ and $c \in \mathcal{G}$.

In Appendix~\ref{Sec_RepTh}, we show that the minimal-dimension representation of this algebra has dimension $M_n = \abs{\mathcal{C}_x}^{n-1}$, where $\abs{\mathcal{C}_x}$ is the number of anyons in $\mathcal{C}_x$. Thus, we obtain
\begin{equation}
  \tmi = \ln \sqrt{\abs{\mathcal{C}_x}}.
\end{equation}

The above discussion straightforwardly generalizes to the case in which a subgroup of anyons with multiple generators is proliferated.

\subsection{Example: decohered toric code phase}

As a concrete example, we compute TEN and TMI of the decohered $\Z_N$ toric code phase. The $\Z_N$ toric code is a paradigmatic Abelian TO in two dimensions. As a straightforward generalization of the $\Z_2$ toric code, it supports a richer spectrum of anyonic excitations described by the group $\Z_N \times \Z_N$, including electric ($e$), magnetic ($m$), and composite quasiparticles of the form $e^a m^b$.

Mixed states obtained by decohering the toric code phase have been studied extensively~\cite{Fan:2023rvp, Ellison:2024svg, wang2025intrinsic}. For a detailed discussion of such decohered states realized on the lattice, see Ref.~\cite{Ellison:2024svg}.

We consider the case in which the anyon $x = e^c m$ is proliferated, with $c \in \hkh{0, 1, \cdots, N-1}$. The subgroup of anyons that braid trivially with $x$ is
\begin{equation}
  \mathcal{C}_x = \set{ e^{-bc}m^b | b=0, 1, \cdots, N-1 } \cong \Z_N,
\end{equation}
which immediately yields
\begin{equation}
  \tmi = \ln \sqrt{N}.
\end{equation}

To compute TEN, we need to identify the transparent anyons in $\mathcal{C}_x$. For an anyon $e^{-bc} m^b$ to braid trivially with all other anyons in $\mathcal{C}_x$, the condition
\begin{equation}
  2bc \equiv 0 \pmod{N}
\end{equation}
must be satisfied. It follows that there are $\gcd(N,2c)$ transparent anyons. Hence, the modular part of $\mathcal{C}_x$ has order $N/\gcd(N,2c)$, and therefore
\begin{equation}
  \ten = \ln \sqrt{\frac{N}{\gcd(N,2c)}}.
\end{equation}
We note that the same result has been obtained in Ref.~\cite{MongUnpublished} using a tensor network approach.

\section{\texorpdfstring{TEN and TMI of decohered $G$-graded string-net states}{TEN and TMI of decohered G-graded string-net states}}\label{Sec_GradedStringNet}

Much of the framework developed in Sec.~\ref{Sec_TQFT} extends to decohered non-Abelian TO. However, additional subtleties arise from the richer algebraic structure of non-Abelian theories. The most significant difference concerns the description of topological boundary conditions: in the non-Abelian setting, one must contend with the full complexity of Lagrangian algebras. At present, it is unclear how to construct a modified Lagrangian algebra that incorporates decoherence effects in a manner analogous to Eq.~\eqref{Eq_TQFT_DecoheredBoundary}.

In this section, we study TEN and TMI of a family of models that host non-Abelian anyons yet remain analytically tractable: decohered $G$-graded string-net models. We first analyze the decohered doubled Ising string-net state as a representative example and then generalize to the full class of decohered $G$-graded string-net models.

\subsection{Decohered doubled Ising string-net state}

We begin with a brief review of the doubled Ising string-net state. The model is defined on a honeycomb lattice with a three-dimensional local Hilbert space on each edge, spanned by the string types $\ket{1}$, $\ket{\sigma}$, and $\ket{\psi}$. A string-net configuration is an assignment of a string type to every edge, i.e., $X = \hkh{s_e}_{e \in \textrm{edges}}$ with $s_e \in \hkh{1, \sigma, \psi}$. We denote the corresponding basis vector by
\begin{equation}
  \ket{X} = \bigotimes_{e \in \textrm{edges}} \ket{s_e}.
\end{equation}
Not every assignment is allowed: a configuration is admissible if at every trivalent vertex the incident labels satisfy the Ising branching constraints, namely (1) a $\sigma$ string cannot terminate, and (2) a $\psi$ string can terminate only when a $\sigma$ string passes through the vertex. In what follows, we refer to admissible configurations as (Ising) string-net configurations and their corresponding basis vectors as (Ising) string-net states.

Graphically, the string-net configurations feature $\sigma$ strings forming loops, while $\psi$ strings either form closed loops or terminate on the $\sigma$ strings. For later use, we define the $\Z_2$ grading operator $\mu_e^z$, which acts on edge $e$ as $\mu_e^z \ket{\sigma} = - \ket{\sigma}$ and $\mu_e^z \ket{1/\psi} = \ket{1/\psi}$.

The ground state wavefunction $\ket{\Psi}$ is a superposition of all string-net states. Consider the ground state on a simply connected lattice. For a given string-net state $\ket{X}$ corresponding to a string-net configuration $X$, the normalized amplitude in the wavefunction is given by~\cite{heinrich2016symmetry}
\begin{equation}
  \braket{X|\Psi} = 2^{\frac12 N_\sigma - N} f(X),
\end{equation}
where $N_\sigma$ is the number of $\sigma$ loops in $X$, $N$ is the total number of plaquettes, and $f(X) \in \hkh{0,\pm 1}$. The explicit expression for $f(X)$ is not essential for the present discussion.

It is useful to introduce an alternative representation of the ground state. Let a configuration of $\sigma$ loops be denoted by $\hkh{\sigma}$, and define a projector $P_{\hkh{\sigma}}$ that annihilates any string-net state whose $\sigma$ loops do not match $\hkh{\sigma}$. Using this projector, we define
\begin{equation}
  \ket{\hkh{\sigma}} = \sum_{X_{\hkh{\sigma}}} 2^{(N_\sigma-N)/2} f(X_{\hkh{\sigma}}) \ket{X_{\hkh{\sigma}}},
\end{equation}
where $X_{\hkh{\sigma}}$ denotes a string-net configuration whose $\sigma$ loop configuration matches $\hkh{\sigma}$. The ground state wavefunction of the Ising string-net model can then be expressed as
\begin{equation}
  \ket{\Psi} = \frac{1}{2^{N/2}} \sum_{\hkh{\sigma}} \ket{\hkh{\sigma}}.
\end{equation}

It is worth noting that, in the absence of any $\sigma$ loops, the state $\ket{\hkh{\sigma} = \varnothing}$ is just a quantum superposition of all closed $\psi$ loops, which is equivalent to a $\Z_2$ toric code state. The introduction of $\sigma$ loops can be interpreted as the insertion of certain topological defect loops into the toric code. In fact, the doubled Ising TO can be obtained from the toric code by gauging the $\Z_2$ symmetry.

We now consider the effect of the following decoherence channel~\cite{Ellison:2024svg}:
\begin{equation}\label{Eq_SN_Ising_Noise}
  \mathcal{N} = \prod_e \mathcal{N}_e, \quad
  \mathcal{N}_e(\rho) = \frac12 (\rho + \mu_e^z \rho \mu_e^z).
\end{equation}
Applying this channel to the pure ground state $\rho_0 = \op{\Psi}$ produces the mixed state
\begin{equation}
  \rho = \mathcal{N}(\rho_0) = \frac{1}{2^N} \sum_{\hkh{\sigma}} \op{\hkh{\sigma}} = \frac{1}{2^N} \sum_{\hkh{\sigma}} \rho_{\hkh{\sigma}},
\end{equation}
where $\rho_{\hkh{\sigma}} = \op{\hkh{\sigma}}$. In this ensemble, the $\sigma$ loops only proliferate probabilistically, while the $\psi$ lines can still fluctuate coherently for fixed $\sigma$ loops.

The Ising string-net state supports nine anyon types labeled by $a \bar{b}$, where $a, b \in \hkh{1, \sigma, \psi}$, and each component obeys the Ising fusion rules. This is known as the doubled Ising TO.

The decoherence channel in Eq.~\eqref{Eq_SN_Ising_Noise} proliferates $\psi \bar{\psi}$. The resulting mixed state $\rho$ is characterized by the anyons that braid trivially with $\psi \bar{\psi}$, namely $1 \bar{1}$, $1 \bar{\psi}$, $\psi \bar{1}$, $\psi \bar{\psi}$, and $\sigma \bar{\sigma}$. Equivalently, these are the excitations whose string operators preserve the $\sigma$ loop configuration. Among them, $\psi \bar{\psi}$ is a nontrivial transparent anyon.

If we further condense $\psi \bar{\psi}$, the anyons $1 \bar{\psi}$ and $\psi \bar{1}$ become identified and $\sigma \bar{\sigma}$ has to split into two Abelian anyons. The theory is reduced to the anyon theory of the $\Z_2$ toric code, with the following relations:
\begin{equation}
  \begin{split}
    \sigma \bar{\sigma} \sim e + m, \\
    1 \bar{\psi} \cong \psi \bar{1} \sim em.
  \end{split}
\end{equation}

To compute TEN for the mixed state $\rho$, we partition the system into $A$ and $\bar{A}$ and consider the partial transpose $\rho^{\intercal_A}$. Using the linearity of the partial transpose, we obtain
\begin{equation}
  \rho^{\intercal_A} = \frac{1}{2^N} \sum_{\hkh{\sigma}} \rho_{\hkh{\sigma}}^{\intercal_A}.
\end{equation}
Crucially, the states $\rho_{\hkh{\sigma}}^{\intercal_A}$ have mutually orthogonal support, as each corresponds to a distinct $\sigma$ loop configuration. As a result, $\rho^{\intercal_A}$ is block-diagonal, and the logarithmic negativity of $\rho$ is given by the sum of the logarithmic negativities of the individual $\rho_{\hkh{\sigma}}$. Since $\rho_{\hkh{\sigma}}$ is a pure state, its logarithmic negativity is the same as the R\'enyi entropy of order $1/2$.

Following Ref.~\cite{levin2006detecting}, the reduced density matrix $\Tr_{\bar{A}} \rho_{\hkh{\sigma}}$ for a simply connected region $A$ takes the form
\begin{equation}\label{Eq_SN_Ising_Reduced_Density_Matrix}
  \Tr_{\bar{A}} \rho_{\hkh{\sigma}} = \frac{1}{2^{L-1-j_{\sigma}}} \sum_{l} \op{\psi_{\hkh{\sigma}}(l)},
\end{equation}
where $L$ denotes the number of edges along the boundary of $A$, $j_{\sigma}\in\mathds{N}$ is defined as half the number of $\sigma$ strings intersecting the boundary of $A$, and $\ket{\psi_{\hkh{\sigma}}(l)}$ are orthonormal states inside $A$ labeled by $l$. The logarithmic negativity is therefore given by
\begin{equation}
  \mathcal{E}_A(\rho_{\hkh{\sigma}}) = (L - 1 - j_{\sigma}) \ln 2.
\end{equation}

Summing over all possible $\sigma$ loop configurations, we obtain
\begin{equation}
  \mathcal{E}_A(\rho) = \frac{1}{2^{L-1}} \sum_{j_{\sigma}=0}^{\lfloor L/2 \rfloor} \binom{L}{2j_{\sigma}} \mathcal{E}_A(\rho_{\hkh{\sigma}}) = \frac{3\ln 2}{4} L -\ln 2.
\end{equation}
Therefore, $\ten = \ln 2$, matching the value in the pure $\Z_2$ toric code. As in the Abelian examples discussed earlier, TEN continues to capture the modular part of the anyon theory of the decohered state.

Because the partial trace is also linear, TMI can be evaluated in a similar manner using Eq.~\eqref{Eq_SN_Ising_Reduced_Density_Matrix}. We postpone the general derivation to Sec.~\ref{Sec_GradedSN} and Appendix~\ref{Sec_TENandTMIofStringNet}. The final result is $\tmi = \frac{3}{2} \ln 2$, which captures the full anyon theory of the decohered state.

\subsection{\texorpdfstring{Decohered $G$-graded string-net states}{Decohered G-graded string-net states}}\label{Sec_GradedSN}

Building on the preceding section, we extend the analysis of the decohered Ising string-net state to the more general case of decohered $G$-graded string-net states first studied in Ref.~\cite{Ellison:2024svg}.

A general string-net model is defined using a unitary fusion category $\mathcal{C}$ as input. The local Hilbert space on each edge is spanned by the simple objects of $\mathcal{C}$ (string types), and the branching rules at vertices are determined by the fusion rules of $\mathcal{C}$. The resulting ground state of the string-net model realizes the TO described by the Drinfeld center $\mathcal{Z}(\mathcal{C})$.

$G$-graded string-net models arise when the input category $\mathcal{C}_G$ is $G$-graded, where $G$ is a finite group. $\mathcal{C}_G$ admits a decomposition of the form
\begin{equation}
  \mathcal{C}_G = \bigoplus_{\mathbf{g} \in G} \mathcal{C}_{\mathbf{g}},
\end{equation}
such that the fusion rules respect the $G$-grading structure, i.e., $\mathcal{C}_{\mathbf{g}} \times \mathcal{C}_{\mathbf{h}} \subset \mathcal{C}_{\mathbf{gh}}$.

Consider the subcategory $\mathcal{C}_{\mathbf{1}} \subset \mathcal{C}_G$, which corresponds to the identity element of the group $G$. In this language, $\mathcal{C}_G$ is a $G$-extension of $\mathcal{C}_{\mathbf{1}}$. The TO $\mathcal{Z}(\mathcal{C}_G)$ can be obtained from $\mathcal{Z}(\mathcal{C}_{\mathbf{1}})$ by gauging the $G$ symmetry; conversely, condensing the $G$ gauge charges, described by $\Rep(G)$, in $\mathcal{Z}(\mathcal{C}_G)$ returns $\mathcal{Z}(\mathcal{C}_{\mathbf{1}})$.

The previously discussed doubled Ising string-net model fits naturally into this framework. The Ising fusion category admits $\Z_2 = \hkh{{\mathbf{1}}, {\mathbf{g}}}$ grading, with $\mathcal{C}_{\mathbf{1}} = \hkh{1, \psi}$ and $\mathcal{C}_{\mathbf{g}} = \hkh{\sigma}$.

The decoherence channel considered earlier proliferates $\Z_2$ gauge charges. A natural generalization is to define the decoherence channel for a $G$-graded string-net model as follows:
\begin{equation}\label{Eq_SN_Noise}
  \mathcal{N} = \prod_e \mathcal{N}_e, \quad
  \mathcal{N}_e(\rho) = \frac{1}{\abs{G}} \sum_{\mathbf{g} \in G} T_e^{\mathbf{g}} \rho T_e^{\mathbf{g}},
\end{equation}
where the operator $T_e^{\mathbf{g}}$ acts on edge $e$ by
\begin{equation}
  T^{\mathbf{g}}_e \ket{a_{\mathbf{h}}} = \delta_{\mathbf{g},\mathbf{h}} \ket{a_{\mathbf{h}}},
\end{equation}
for $\mathbf{g}, \mathbf{h} \in G$ and $a_{\mathbf{h}} \in \mathcal{C}_{\mathbf{h}}$.

To study the decohered ground state, we introduce the notion of a $G$-defect network, denoted by $\hkh{\mathbf{g}}$, which assigns a group element to each edge. This assignment must satisfy group multiplication constraints at each vertex. The concept is analogous to the configuration of $\sigma$ loops $\hkh{\sigma}$ in the doubled Ising case. We also define a projector $P_{\hkh{\mathbf{g}}}$, which projects to string-net states compatible with a given $G$-defect network.

Applying the channel in Eq.~\eqref{Eq_SN_Noise} to the ground state $\rho_0=\op{\Psi}$ yields
\begin{equation}\label{Eq_SN_rho}
  \rho = \mathcal{N}(\rho_0) = \frac{1}{\abs{G}^N} \sum_{\hkh{\mathbf{g}}} \op{\Psi_{\hkh{\mathbf{g}}}} = \frac{1}{\abs{G}^N} \sum_{\hkh{\mathbf{g}}} \rho_{\hkh{\mathbf{g}}},
\end{equation}
where $N$ is the total number of plaquettes and $\ket{\Psi_{\hkh{\mathbf{g}}}} = P_{\hkh{\mathbf{g}}} \ket{\Psi}$. Physically, the operators $T_e^{\mathbf{g}}$ create $G$ gauge charges and the decohered state $\rho$ is obtained from $\rho_0$ by proliferating these gauge charges. Notably, when $G$ is a non-Abelian group, the gauge charges may contain non-Abelian anyons.

We now summarize TEN and TMI for $\rho$; the full derivations are provided in Appendix~\ref{Sec_TENandTMIofStringNet}.

For TEN, $\rho^{\intercal_A}$ is again block-diagonal, and we only need to sum the R\'enyi entropy $S_A^{(\frac12)}(\rho_{\hkh{\mathbf{g}}})$ over all possible $G$-defect networks. For each $G$-defect network $\hkh{\mathbf{g}}$, the reduced density matrix $\Tr_{\bar{A}} \rho_{\hkh{\mathbf{g}}}$ is diagonal; unlike the doubled Ising case, however, it is not generally an equal-weight mixture, which complicates the evaluation of the R\'enyi entropy.

We establish explicit results for TEN in two scenarios: (1) when the fusion category $\mathcal{C}_{\mathbf{1}}$ is Abelian, and (2) when the group $G$ is Abelian. In both cases, we find that TEN is given by
\begin{equation}
  \ten = \ln \cD_{\mathbf{1}}^2.
\end{equation}

For TMI in this lattice model, we use Eq.~\eqref{Eq_EM_CMI} together with the Levin--Wen geometry in Fig.~\ref{Fig_TMIAnnulus}. For a simply connected region $A$, the entanglement entropy is given by
\begin{equation}
  \begin{split}
    S_A(\rho) = N_A \ln \abs{G} + & \fkh{\ln (\abs{G} \cD_{\mathbf{1}}^2) - \frac{1}{\abs{G} \cD_{\mathbf{1}}^2} \sum_{\mathbf{g}} N_{\mathbf{g}}} L\\
    & - \ln (\abs{G} \cD_{\mathbf{1}}^2),
  \end{split}
\end{equation}
where $N_A$ is the number of plaquettes in $A$, and the first term is the expected volume-law contribution. Combining the entropies according to Eq.~\eqref{Eq_EM_CMI} yields
\begin{equation}
  \tmi = \ln(\sqrt{\abs{G}} \cD_{\mathbf{1}}^2).
\end{equation}

\section{One-form symmetry of mixed-state TO}\label{Sec_1form}

For ground-state topological states, TEE is related to the total quantum dimension $\cD$ of the anyon theory. In fact, it is now understood that $\ln \cD$ gives the universal (tight) lower bound for $\gamma$.

One way to understand anyon theory just from the ground state is to view it as the theory of (emergent) one-form symmetries in the ground state, including both invertible and non-invertible ones. Recent work proposes that mixed-state TOs in two dimensions can be similarly characterized by their (generally emergent) \emph{strong} 1-form symmetries. We briefly review the approach here.

First, we recall the notion of strong and weak symmetry. For a mixed state $\rho$, a unitary operator $U$ is a strong symmetry if
\begin{equation}\label{Eq_1form_Strong_Sym}
  U \rho = \euler^{\ii\theta} \rho,
\end{equation}
for some phase factor $\euler^{\ii\theta}$. It implies that in any pure state decomposition of $\rho$, all pure states are eigenstates of $U$ with the same eigenvalue $\euler^{\ii\theta}$. If $U$ does not satisfy Eq.~\eqref{Eq_1form_Strong_Sym} but instead
\begin{equation}
  U \rho U^\dagger = \rho,
\end{equation}
then $U$ is called a weak symmetry.

The notion of strong symmetry can be easily generalized to non-invertible operators, in which case the eigenvalue is no longer a phase factor.

In two dimensions, a strong 1-form symmetry is a strong symmetry generated by closed string operators. It is widely accepted that ground-state TOs are completely characterized by the (emergent) 1-form symmetries (up to invertible states), which can be thought of as string operators for anyon excitations. Mathematically, the string operators are described by a unitary MTC. Here, modularity is required by the physical principle of remote detectability.

Decohered topological phases can also be associated with strong 1-form symmetries. Its importance is established by the following result: strong symmetries can be ``pushed through'' a quasi-local quantum channel. Specifically, if $\rho_1 = \mathcal{N}(\rho_2)$ for a quasi-local channel $\mathcal{N}$, then, upon purification of the channel, the state $\rho_2$ shares all strong 1-form symmetries of $\rho_1$, with the categorical data related in a definite way as described in Ref.~\cite{Ellison:2024svg}. A similar observation is made in Ref.~\cite{Lessa:2025xut} and is termed ``symmetry pullback''. A crucial difference from the ground-state case is that the relevant category needs only to be premodular rather than modular \cite{Ellison:2024svg, Sohal:2024qvq, wang2025intrinsic}. Two mixed-state TOs characterized by distinct premodular categories must belong to different phases; that is, they cannot be connected by quasi-local channels in both directions.

We are mainly interested in mixed-state TOs obtained from decohering ground-state TOs. Denote the anyon theory of the ground-state TO by $\mathcal{C}$. The decohered TO has a different anyon theory, denoted by $\mathcal{C}'$. In most of our examples, $\mathcal{C}'$ is a subcategory of $\mathcal{C}$, although this is not true in general. As a premodular category, $\mathcal{C}'$ has a transparent subcategory $\mathcal{T}$. Condensing $\mathcal{T}$ yields another modular anyon theory, denoted by $\mathcal{M}$.

Our results (both the replica field-theory calculations for decohered Abelian TOs, and the exact results obtained for decohered $G$-graded string-net models) can be summarized as follows:
\begin{equation}
  \ten = \ln \cD_{\mathcal{M}},
\end{equation}
and
\begin{equation}
  \tmi = \ln \cD_{\mathcal{C}'}.
\end{equation}
Here, $\cD$ denotes the total quantum dimension of the corresponding anyon theory.

That is to say, TMI captures the total quantum dimension of the anyon theory $\mathcal{C}'$ derived from the strong 1-form symmetry of the mixed-state TO, while the ``universal part'' of TEN captures the modular part $\mathcal{M}$.

Just as in the case of TEE, it is generally expected that there can be ``spurious'' contributions. Indeed, Ref.~\cite{Lessa:2025xut} proved that for Abelian theories $\ln \cD_{\mathcal{C}'}$ is a lower bound for $\tmi$. As already discussed in Sec.~\ref{Sec_TQFT}, our field-theoretical calculations assume that no additional (symmetry-protected or accidental) degeneracy occurs on the defects in the boundary theory. So strictly speaking, the results in Sec.~\ref{Sec_TQFT} should also be understood as lower bounds for $\ten$ and $\tmi$. On the other hand, the microscopic calculations in Sec.~\ref{Sec_GradedStringNet} are exact and hence are free of any ``spurious'' contributions for either $\ten$ or $\tmi$.

\section{Conclusions and discussions}

In this work, we investigated TEN and TMI as diagnostic tools for mixed-state TOs emerging from decohered pure states. By analyzing these quantities, we established a correspondence between the entanglement data and the underlying anyon structure of the decohered phase. For Abelian TOs, we developed a replica field-theoretic framework that maps the computation of TEN and TMI to the quantum dimensions of domain-wall defects separating decoherence-induced topological boundary conditions.

Our findings suggest that distinct physical information is captured by these two measures. Guided by the strong one-form symmetries that characterize the mixed-state phase, we showed that TMI probes the total quantum dimension of the resulting emergent premodular anyon theory. In contrast, TEN acts as a sharper filter, detecting only the modular sector of this theory. We further corroborated this physical picture through exact calculations in decohered G-graded string-net models, extending the validity of our conclusions to systems with non-Abelian anyons.

We now discuss open questions.

First, recent work on the $\Z_2$ toric code stabilizer model under fermion decoherence has found that the negativity remains invariant regardless of decoherence strength~\cite{wang2025intrinsic}; consequently, TEN fails to capture the phase transition in this example. Furthermore, depending on the choice of entanglement cut, TEN exhibits a lattice-scale sensitivity to the parity of the region length. These non-universal features stand in contrast to our field-theoretic prediction for the $\Z_2$ toric code model with $em$ proliferation, which predict $\ten=0$ in the strong decoherence limit. While this persistent negativity can be explained by the correlated nature of the decoherence channel~\cite{kim2024persistent}, a comprehensive understanding of TEN's dependence on microscopic details is still lacking. An important open question is whether the universal values derived in our framework can be established as bounds for generic cases.

Second, the analysis in this paper is sharpest in the strong-decoherence regime, leaving the behavior at finite decoherence strength less understood: how the universal terms cross over, and what additional nonuniversal contributions can obscure their extraction.

Finally, we address the current limitations of our field-theoretic framework regarding non-Abelian topological orders. While the general approach extends to the non-Abelian setting, additional subtleties arise from the more complicated algebraic structure of these theories. The most significant challenge lies in the description of topological boundary conditions, which involves the full complexity of Lagrangian algebras. Resolving this issue is essential for a complete field-theoretic description.

\section{Acknowledgment}

We are grateful to Tyler Ellison, Carolyn Zhang, and Roger Mong for many discussions and for their participation at the early stage of this project. MC would like to thank Sagar Vijay for enlightening conversations and Tyler Ellison for comments on a draft of the manuscript. MC and KLC are partially supported by NSF Grant No.~DMR-2424315. MC is grateful to the Institute for Advanced Study for its hospitality.

\bibliography{ref.bib}

\onecolumngrid

\appendix

\section{Representation theory of string operator algebra}\label{Sec_RepTh}

As demonstrated in Sec.~\ref{Sec_TQFT}, the calculation of TEN and TMI of (decohered) Abelian TO reduces to determining the minimal-dimension representation of the string operator algebra generated by $T(j,b)$, which satisfy the commutation relations
\begin{equation}\label{Eq_RepTh_Algebra}
  T(j,b) T(j+1,c) = B(b,c) T(j+1,c) T(j,b),
\end{equation}
where $B(b,c)$ is the braiding phase between anyons $b$ and $c$. The range of the index $j$ and anyon label $b$ in $T(j,b)$ are summarized as follows:
\begin{itemize}
  \item Pure state TO:
    \begin{itemize}
      \item TEN: $j=1, 2, \cdots, n$ and $b \in \mathcal{G}$.
      \item TMI: $j=1, 2, \cdots, 2n$ and $b \in \mathcal{G}$.
    \end{itemize}
  \item Decohered TO:
    \begin{itemize}
      \item TEN: $j=1, 2, \cdots, n$ and $b \in \mathcal{G}_x$.
      \item TMI: $j=1, 2, \cdots, 2n$, $b \in \mathcal{G}_x$ for $j$ odd and $b \in \mathcal{G}$ for $j$ even.
    \end{itemize}
\end{itemize}
In the case of TEN, the number of replicas $n$ must be even. Consequently, the upper limit of $j$ is always an even integer. For notational convenience, we will henceforth write $j=1, 2, \cdots, m$, where $m$ is understood to be even, $m=n$ for TEN and $m=2n$ for TMI.

We address this problem by showing that the algebra Eq.~\eqref{Eq_RepTh_Algebra} is equivalent to multiple copies of the torus string operator algebra.

First, recall the structure of the torus string operator algebra. Let $\gamma_1$ and $\gamma_2$ represent the two non-contractible loops of a torus, and let $T_{\gamma_i}(b)$ denote the Wilson loop operator corresponding to anyon $b$ along loop $\gamma_i$. The torus string operator algebra satisfies the commutation relations
\begin{equation}\label{Eq_RepTh_TorusAlgebra}
  T_{\gamma_1}(b) T_{\gamma_2}(c) = B(b,c) T_{\gamma_2}(c) T_{\gamma_1}(b).
\end{equation}
For pure state TO, the minimal-dimension representation of the torus string operator algebra is the ground state space of the TO on a torus, which has dimension $\cD^2$.

To simplify the commutation relations Eq.~\eqref{Eq_RepTh_Algebra}, we perform a change of basis by introducing a new set of generators $P(j,b)$ defined as
\begin{equation}\label{Eq_RepTh_BasisTrans}
  P(j,b) =
  \begin{cases}
    T(1,b) T(3,b) \cdots T(j,b) & \quad j \textrm{ odd} \\[7pt]
    T(j,b)                      & \quad j \textrm{ even, } j \neq m \\[7pt]
    T(2,b) T(4,b) \cdots T(m,b) & \quad j=m
  \end{cases}.
\end{equation}
In terms of these new generators, the commutation relations take the form
\begin{equation}
  P(2j-1,b) P(2j,c) = B(b,c) P(2j,c) P(2j-1,b), \quad j=1, 2, \cdots, m/2-1,
\end{equation}
which is precisely $m/2-1$ independent copies of the torus string operator algebra Eq.~\eqref{Eq_RepTh_TorusAlgebra}. As a result, the minimal-dimension representation of the original algebra can be constructed by taking the tensor product of the minimal-dimension representations of each torus algebra copy. In the case of pure state TO, this implies that the smallest dimension is $\cD^{m-2}$.

We now analyze the minimal-dimension representation of the torus string operator algebra Eq.~\eqref{Eq_RepTh_TorusAlgebra} in the context of decohered TO. To compute TEN, the relevant Wilson loop operators $T_{\gamma_1}(b)$ and $T_{\gamma_2}(c)$ are restricted to $b, c \in \mathcal{G}_x$. This setup corresponds to the string operator algebra of the anyon theory $\mathcal{C}_x$ on a torus. The minimal-dimension representation of this algebra has dimension $\cD_{\mathcal{M}_x}^2 = \abs{\mathcal{M}_x}$, where $\mathcal{M}_x$ is the modular part of $\mathcal{C}_x$ and $\cD_{\mathcal{M}_x}$ is its total quantum dimension. This is because transparent anyons do not contribute to nontrivial commutation relations and hence do not enlarge the representation space. Additional details supporting this claim, as well as the motivation for the basis transformation introduced in Eq.~\eqref{Eq_RepTh_BasisTrans}, are provided in Appendix~\ref{Sec_RepTh_TwistedApplication}.

For the computation of TMI, we have $b \in \mathcal{G}_x$ for $T_{\gamma_1}(b)$ and $c \in \mathcal{G}$ for $T_{\gamma_2}(c)$. The fact that $c$ takes value in $\mathcal{G}$ rather than $\mathcal{G}_x$ implies that the minimal-dimension representation of the algebra has a dimension larger than $\cD_x^2$. This indicates that TMI captures more than just the modular part $\mathcal{M}_x$. Since we started with a modular theory, the transparent anyons in $\mathcal{C}_x$ necessarily braid nontrivially with at least one anyon in $\mathcal{C}$. In the following, we will show that the minimal dimension is $\cD_{\mathcal{C}_x}^2 = \abs{\mathcal{C}_x}$.

We work in a basis where the operators $T_{\gamma_2}(c)$ are diagonal. Let $\ket{a}$ be a state with $T_{\gamma_2}(c) \ket{a} = \lambda(c) \ket{a}$. For $b \in \mathcal{C}_x$, define $\ket{b} = T_{\gamma_1}(b) \ket{a}$. Using the commutation relations of the torus algebra, it follows that
\begin{equation}
  T_{\gamma_2}(c) \ket{b} = B(b,c)^{-1} \lambda(c) \ket{b}.
\end{equation}
Using the unitarity of $T_{\gamma_2}(c)$, we compute the inner product between two such states:
\begin{equation}
  \braket{b_1|b_2} = \braket{b_1 | T_{\gamma_2}^{-1}(c) T_{\gamma_2}(c) | b_2} = B(b_2 \bar{b}_1, c)^{-1} \braket{b_1|b_2}.
\end{equation}
As long as $b_1 \neq b_2$, by modularity there exists $c \in \mathcal{C}$ such that $B(b_2 \bar{b}_1, c) \neq 1$. Therefore, we find
\begin{equation}
  \braket{b_1|b_2} = \delta_{b_1,b_2}.
\end{equation}
Thus, the states $\hkh{\ket{b}}$ are mutually orthonormal and therefore linearly independent. It follows that the dimension of the representation is at least $\abs{\mathcal{C}_x}$.

Next, we explicitly construct a $\abs{\mathcal{C}_x}$-dimensional representation that saturates the lower bound. Consider an orthonormal basis $\hkh{\ket{a}}$ labeled by anyons $a \in \mathcal{C}_x$. Define the action of the Wilson loop operators as follows:
\begin{equation}
  \begin{split}
    T_{\gamma_1}(b) \ket{a} = & \ket{b\times a}, \quad b \in \mathcal{G}_x\\
    T_{\gamma_2}(c) \ket{a} = & B(c,a) \ket{a}, \quad c \in \mathcal{G}
  \end{split}\quad .
\end{equation}
It is straightforward to verify the commutation relations are satisfied. Hence, this construction yields a representation of dimension $\abs{\mathcal{C}_x}$.

\subsection{Representation theory of twisted algebra}\label{Sec_RepTh_Twisted}

In this section, we study the representation theory of twisted algebra in general. In particular, we consider the algebra $A$ generated by a finite set of elements $\hkh{T_i}_{i=1}^k$, which satisfy the commutation relations:
\begin{equation}
  T_i T_j = \euler^{4\pi\ii M_{ij}} T_j T_i,
\end{equation}
where $M$ is a $k\times k$ antisymmetric matrix with rational entries.

To find the minimal-dimension representation of $A$, we consider the related Abelian group $G$ generated by $\hkh{w_i}_{i=1}^k$. The key fact is that representations of $A$ correspond to projective representations of $G$, with a factor set $\omega \in H^2(G,U(1))$ determined by $M$. Specifically, the factor set $\omega$ is given by
\begin{equation}
  \omega(w_1^{m_1} \cdots w_k^{m_k}, w_1^{n_1} \cdots w_k^{n_k}) = \exp\kh{2\pi\ii \sum_{ij} m_i M_{ij}n_j}.
\end{equation}
Notably, any factor set can be brought into this form by making appropriate gauge transformations.

Let $\rho: G \to \mathrm{GL}(n,\C)$ be a projective representation corresponding to $\omega$. Consider the image of generators $\rho(w_i)$; they satisfy the relations:
\begin{equation}
  \rho(w_i) \rho(w_j) = \euler^{4\pi\ii M_{ij}} \rho(w_j) \rho(w_i).
\end{equation}
This is precisely the relation satisfied by the generators of the algebra $A$. Therefore, $\rho$ is also a representation of $A$.

Conversely, if we have a representation of the algebra $\rho: A \to \mathrm{GL}(n,\C)$, we obtain a projective representation of $G$ by
\begin{equation}
  \rho(w_1^{m_1} \cdots w_k^{m_k}) = \exp\kh{-2\pi\ii \sum_{i} \sum_{j>i} n_i M_{ij} n_j} \rho^{m_1}(T_1) \cdots \rho^{m_k}(T_k).
\end{equation}
Thus, we have proven that representations of $A$ are in one-to-one correspondence with projective representations of $G$.

According to Ref.~\cite{backhouse1972projective}, finite-dimensional unitary irreducible projective representations of the finitely generated Abelian group $G$ with factor set $\omega$ share a common dimension, denoted as $d(\omega)$, which can be determined as follows.

First, we express the entries of $M$ as reduced fractions: $M_{ij} = \frac{t_{ij}}{2 N_{ij}}$, where $t_{ij}=-t_{ji}$ and $N_{ij}=N_{ji}$. Let $N = \textrm{lcm}(N_{ij})$ be the least common multiple of the $N_{ij}$, and define the matrix $P = 2 N M$. By construction, $P$ is an antisymmetric integer matrix.

The next step is to put the matrix $P$ in its canonical form by congruence. This amounts to making basis transformations of $G$ to simplify $\omega$. There exists a unimodular integer matrix $U$ such that $U^T P U$ is zero except for a certain number $r$ of $2\times 2$ integer anti-symmetric matrices $E_i$ along the diagonal. Furthermore, it is possible to choose the matrices $E_i$ such that the nonzero entries of $E_i$ divide the nonzero entries of $E_{i+1}$. In the following, we use $\sim$ to denote congruence between integer matrices, and
\begin{equation}
  P \sim U^T P U = \bigoplus_{i=1}^r
  \begin{pmatrix}
    0 & a_i\\
    -a_i & 0
  \end{pmatrix} \oplus \vb{0}_{k-2r}
\end{equation}

From the string operator algebra perspective, the original algebra splits into several Pauli algebras. And the dimension of irreducible projective representation $d(\omega)$ is given by
\begin{equation}
  d(\omega) = \prod_{i=1}^r \frac{N}{\gcd(N,a_i)}.
\end{equation}

\subsection{Application to string operator algebra}\label{Sec_RepTh_TwistedApplication}

Here, we apply the theoretical framework developed above to analyze the string operator algebra given in Eq.~\eqref{Eq_RepTh_Algebra}. The corresponding matrices are constructed from the braiding phase $B(b,c) = \euler^{2\pi\ii t_{bc} / N_{bc}}$. Let $N = \textrm{lcm}(N_{bc})$, we define the matrix $P_1$ as $(P_1)_{bc} = N t_{bc} / N_{bc}$. The matrix $P_3$, associated with the string operator algebra Eq.~\eqref{Eq_RepTh_Algebra}, is given by
\begin{equation}
  P_3 =
  \begin{pmatrix}
    \vb{0} & P_1 & \vb{0} & \vb{0} &\cdots &\vb{0} & -P_1\\
    -P_1^{\intercal} & \vb{0} & P_1^{\intercal} & \vb{0} &\cdots &\vb{0} & \vb{0}\\
    \vb{0} & -P_1 & \vb{0} & P_1 &\cdots &\vb{0} & \vb{0}\\
    \vb{0} & \vb{0} & -P_1^{\intercal} & \vb{0} &\cdots &\vb{0} & \vb{0}\\
    \vdots & \vdots & \vdots & \vdots & \ddots & \vdots & \vdots\\
    \vb{0} & \vb{0} & \vb{0} & \vb{0} & \cdots & \vb{0} & P_1\\
    P_1^{\intercal} & \vb{0} & \vb{0} & \vb{0} & \cdots & -P_1^{\intercal} & \vb{0}
  \end{pmatrix}.
\end{equation}
By making elementary row and column operations, we can block diagonalize $P_3$ as
\begin{equation}
  P_3 \sim \bigoplus_{i=1}^{m/2-1}
  \begin{pmatrix}
    \vb{0} & P_1\\
    -P_1^{\intercal} & \vb{0}
  \end{pmatrix} \oplus \vb{0}=P_2^{\oplus \frac{m-2}{2}} \oplus \vb{0},
\end{equation}
where each block $P_2$ corresponds to one copy of the torus string operator algebra Eq.~\eqref{Eq_RepTh_TorusAlgebra}. The unimodular matrix used in this congruence translates to the basis transformation we have defined in Eq.~\eqref{Eq_RepTh_BasisTrans}.

To further analyze the dimension of the representation, we bring the matrix $P_2$ into its canonical form. We focus on the case relevant to TEN, in which the matrix $P_1$ is symmetric. In this setting, $P_1$ can be brought into Smith normal form using a unimodular integer matrix $U_1$, such that
\begin{equation}
  P_1 \sim U_1^{\intercal} P_1 U_1 = \textrm{diag}[a_1, \cdots, a_r, 0, \cdots, 0].
\end{equation}
This transformation corresponds to a change of basis among the anyon generators. The nonzero diagonal entries $a_i$ identify generators associated with the modular part of the anyon theory. In contrast, the zero entries correspond to generators of transparent anyons. The canonical form of $P_2$ follows directly from the Smith normal form of $P_1$:
\begin{equation}
  P_2 \sim \bigoplus_{i=1}^r
  \begin{pmatrix}
    0 & a_i\\
    -a_i & 0
  \end{pmatrix} \oplus \vb{0}.
\end{equation}
It follows that the dimension of the irreducible representation only counts the number of anyons in the modular part.

\section{\texorpdfstring{Derivation of TEN and TMI in decohered $G$-graded string-net states}{Derivation of TEN and TMI in decohered G-graded string-net states}}\label{Sec_TENandTMIofStringNet}

In this section, we compute TEN and TMI of the decohered $G$-graded string-net state $\rho$ constructed in Sec.~\ref{Sec_GradedSN}:
\begin{equation}
  \rho = \frac{1}{\abs{G}^N} \sum_{\hkh{\mathbf{g}}} \rho_{\hkh{\mathbf{g}}}.
\end{equation}

In the following, we will require the notion of the fusion multiplicity $N_{ab}^c$ in a fusion category, which quantifies the number of distinct ways in which objects $a$ and $b$ can fuse to produce the object $c$. The fusion rule—which also determines the branching rules in the associated string-net model—can then be expressed as
\begin{equation}
  a \times b = \sum_c N_{ab}^c\; c.
\end{equation}

Since $\rho^{\intercal_A}$ is block-diagonal, the logarithmic negativity $\mathcal{E}_A(\rho)$ can be computed by summing $\mathcal{E}_A(\rho_{\hkh{\mathbf{g}}})$ over all possible $G$-defect networks $\hkh{\mathbf{g}}$. For each $\hkh{\mathbf{g}}$, $\rho_{\hkh{\mathbf{g}}}$ is a pure state, and we have $\mathcal{E}_A(\rho_{\hkh{\mathbf{g}}}) = S_A^{(\frac12)}(\rho_{\hkh{\mathbf{g}}})$. Similarly, the reduced density matrix $\rho_A$ is also block-diagonal, allowing the entanglement entropy $S_A(\rho)$ to be expressed as a sum over contributions from each $\rho_{\hkh{\mathbf{g}}}$ individually. In the following, we first fix a $G$-defect network $\hkh{\mathbf{g}}$ and examine the reduced density matrix of the corresponding pure state $\rho_{\hkh{\mathbf{g}}}$.

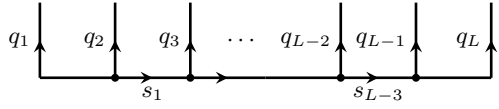
\begin{figure}[ht]
  \centering
  \begin{tikzpicture}[line width=1pt]
    \draw (0,0) -- (1,0);
    \draw (3,0) -- (4,0);
    \draw (5,0) -- (6,0);
    \draw[arrowmid] (0,0) -- (0,1) node[midway, left] {$q_1$};
    \draw[arrowmid] (1,0) -- (1,1) node[midway, left] {$q_2$};
    \draw[arrowmid] (2,0) -- (2,1) node[midway, left] {$q_3$};
    \node at (2.7,0.5) {$\cdots$};
    \draw[arrowmid] (4,0) -- (4,1) node[midway, left] {$q_{L-2}$};
    \draw[arrowmid] (5,0) -- (5,1) node[midway, left] {$q_{L-1}$};
    \draw[arrowmid] (6,0) -- (6,1) node[midway, left] {$q_L$};

    \draw[arrowmid] (1,0) -- (2,0) node[midway, below] {$s_1$};
    \draw[arrowmid] (2,0) -- (3,0);
    \draw[arrowmid] (4,0) -- (5,0) node[midway, below] {$s_{L-3}$};
    \fill (1,0) circle (1.5pt);
    \fill (2,0) circle (1.5pt);
    \fill (4,0) circle (1.5pt);
    \fill (5,0) circle (1.5pt);
  \end{tikzpicture}
  \caption{Reduced tree-like string-net configuration within the subregion $A$. The edges $q_i$ are on the boundary of $A$, while the edges $s_i$ are internal to $A$.}
  \label{Fig_StringNet_InsideA}
\end{figure}

When considering the reduced density matrix $\Tr_{\bar{A}} \rho_{\hkh{\mathbf{g}}}$ for a simply connected region $A$, the string-net configurations within $A$ can be reduced to a minimal, tree-like form along the boundary of $A$ (see Fig.~\ref{Fig_StringNet_InsideA})~\cite{levin2006detecting}. We denote the group element associated with edge $q_i$ by $\mathbf{g}_i$, which are given by $\hkh{\mathbf{g}}$ along the boundary of $A$. Together, the group elements $\hkh{\mathbf{g}_i}$ uniquely specify the remaining $G$-defect network for the tree-like string-net configurations.

The reduced density matrix $\Tr_{\bar{A}} \rho_{\hkh{\mathbf{g}}}$ takes the form:
\begin{equation}\label{Eq_APP_SN_ReducedDensityMatrix}
  \Tr_{\bar{A}} \rho_{\hkh{\mathbf{g}}} \propto \kh{\prod_{m=1}^L d_{q_m}} \op{\hkh{q,s}},
\end{equation}
where $\hkh{q,s}$ denotes a valid tree-like string-net configuration, $\ket{\hkh{q,s}}$ are orthonormal states labeled by $\hkh{q,s}$ and $d_{q_m}$ denotes the quantum dimension of the string $q_m$. A configuration is considered valid if it is compatible with $\hkh{\mathbf{g}_i}$ and satisfies the branching constraints.

The overall normalization of $\Tr_{\bar{A}} \rho_{\hkh{\mathbf{g}}}$ is given by
\begin{align}
  \sum_{\hkh{q,s}} \prod_{m=1}^L d_{q_m}
  = & \sum_{q_1 \in \mathcal{C}_{\mathbf{g}_1}} \cdots \sum_{q_{L-1} \in \mathcal{C}_{\mathbf{g}_{L-1}}} \sum_{q_L} \sum_{s_1, \cdots, s_{L-3}} N^{s_1}_{q_1 q_2} N^{s_2}_{s_1 q_3} \cdots N^{q_L}_{s_{L-3} q_{L-1}} \prod_{m=1}^L d_{q_m}\\
  = & \sum_{q_1 \in \mathcal{C}_{\mathbf{g}_1}} \cdots \sum_{q_{L-1} \in \mathcal{C}_{\mathbf{g}_{L-1}}} \prod_{m=1}^{L-1} d_{q_m}^2 = \prod_{m=1}^{L-1} \kh{\sum_{q_m \in \mathcal{C}_{\mathbf{g}_m}} d_{q_m}^2}\\
  = & \kh{\frac{\cD^2}{\abs{G}}}^{L-1} = \cD_{\mathbf{1}}^{2L-2},
\end{align}
where $\cD$ is the total quantum dimension of the $G$-graded category $\mathcal{C}_G$, $\cD_{\mathbf{1}}$ is the total quantum dimension of the identity sector $\mathcal{C}_{\mathbf{1}}$ and $\abs{G}$ is the order of the group $G$. In the second equality, we applied the identity $\sum_{c} N_{ab}^c d_c = d_a d_b$ successively. In the final equality, we used the fact that the total quantum dimension is equally distributed among the $G$-grading sectors, leading to the relation $\cD^2 = \abs{G} \cD_{\mathbf{1}}^2$.

\subsection{TEN}

The logarithmic negativity of $\rho_{\hkh{\mathbf{g}}}$ is given by
\begin{equation}
  \mathcal{E}_A(\rho_{\hkh{\mathbf{g}}}) = S_A^{(\frac12)}(\rho_{\hkh{\mathbf{g}}}) = 2 \ln \kh{\frac{1}{\cD_{\mathbf{1}}^{L-1}} \sum_{\hkh{q,s}} \prod_{m=1}^L d_{q_m}^{\frac12}}. 
\end{equation}
To proceed, we consider two scenarios: (1) when the fusion category $\mathcal{C}_{\mathbf{1}}$ is Abelian, and (2) when the group $G$ is Abelian.

In the first case, where the fusion category $\mathcal{C}_{\mathbf{1}}$ is Abelian, the quantum dimension $d_{q_m}$ is uniquely determined by the group element $\mathbf{g}_m$. That is, all string types within $\mathcal{C}_{\mathbf{g}_m}$ share the same quantum dimension. Consequently, $\rho_{\hkh{\mathbf{g}}}$ is an equal-weight distribution, and the R\'enyi entropy coincides with the von Neumann entropy, which significantly simplifies the calculation. We find
\begin{equation}
  \mathcal{E}_A(\rho_{\hkh{\mathbf{g}}}) = (L-1) \ln \cD_{\mathbf{1}}^2 - \frac{1}{\cD_{\mathbf{1}}^2} \sum_{m=1}^L N_{\mathbf{g}_m},
\end{equation}
where $N_{\mathbf{g}} = \sum_{a\in\mathcal{C}_{\mathbf{g}}} d_a^2 \ln d_a$.

To find $\mathcal{E}_A(\rho)$, we need to average $\mathcal{E}_A(\rho_{\hkh{\mathbf{g}}})$ over all possible $G$-defect networks. This is equivalent to averaging over all possible $G$-defect networks in the reduced tree-like form (Fig.~\ref{Fig_StringNet_InsideA}). We have
\begin{equation}
  \mathcal{E}_A(\rho)
  =\kh{\ln \cD_{\mathbf{1}}^2 - \frac{1}{\abs{G} \cD_{\mathbf{1}}^2} \sum_{\mathbf{g} \in G} N_{\mathbf{g}}} L - \ln \cD_{\mathbf{1}}^2.
\end{equation}
In conclusion, TEN is given by
\begin{equation}
  \ten = \ln \cD_{\mathbf{1}}^2,
\end{equation}
which is equal to that of a pure TO described by $\mathcal{Z}(\mathcal{C}_{\mathbf{1}})$.

In the second case, where the group $G$ is Abelian, the leading contribution to the R\'enyi entropy can be evaluated using a method analogous to that in Ref.~\cite{flammia2009topological}. To this end, we introduce a vector space with an orthonormal basis $\hkh{\ket{a}}$, where the labels $a \in \mathcal{C}_G$ correspond to the string types. We define a set of matrices $N_q$ by
\begin{equation}
  N_{q} = \sum_{a,b} N_{aq}^b \ket{a} \bra{b},
\end{equation}
where $N_{aq}^b$ are the fusion multiplicities.

The summation over valid string-net configurations can be reformulated as
\begin{align}
  \sum_{\hkh{q,s}} \prod_{m=1}^L d_{q_m}^{\frac12}
  = & \sum_{q_1 \in \mathcal{C}_{\mathbf{g}_1}} \cdots \sum_{q_L \in \mathcal{C}_{\mathbf{g}_L}} \braket{1| \prod_{m=1}^L N_{q_m} d_{q_m}^{\frac12} |1} \\
  = & \braket{1| \prod_{m=1}^L\; \sum_{q_m \in \mathcal{C}_{\mathbf{g}_m}} N_{q_m} d_{q_m}^{\frac12} |1} \\
  = & \braket{1| \prod_{m=1}^L\; M_{\mathbf{g}_m} |1},
\end{align}
where $\ket{1}$ is the basis vector corresponding to the trivial string, and $M_{\mathbf{g}} = \sum_{q \in \mathcal{C}_{\mathbf{g}}} N_q d_q^{\frac12}$. In the limit of large $L$, the leading contribution arises from the leading eigenvalues of $M_{\mathbf{g}}$ (with the largest magnitude) and their corresponding eigenvectors.

It is known that the quantum dimension $d_q$ is a leading eigenvalue of $N_q$, and the corresponding eigenvector is
\begin{equation}
  \ket{d} = \frac{1}{\cD} \sum_{a \in \mathcal{C}_G} d_a \ket{a} = \frac{1}{\cD} \sum_{\mathbf{h} \in G} \sum_{a \in \mathcal{C}_{\mathbf{h}}} d_a \ket{a}.
\end{equation}
Therefore, $\ket{d}$ is also an eigenvector of $M_{\mathbf{g}}$ associated with the leading eigenvalue $\lambda(\mathbf{g}) = \sum_{q \in \mathcal{C}_{\mathbf{g}}} d_q^{\frac32}$. However, it is not unique. For any character $\chi_k$ of the finite Abelian group $G$, the following holds for $q \in \mathcal{C}_{\mathbf{g}}$:
\begin{equation}
  N_q \kh{\frac{1}{\cD} \sum_{\mathbf{h} \in G} \sum_{a \in \mathcal{C}_{\mathbf{h}}} \chi_k(\mathbf{h}) d_a \ket{a}} = \chi_k^{-1}(\mathbf{g}) d_q \kh{\frac{1}{\cD} \sum_{\mathbf{h} \in G} \sum_{a \in \mathcal{C}_{\mathbf{h}}} \chi_k(\mathbf{h}) d_a \ket{a}}.
\end{equation}
There are a total of $\abs{G}$ such eigenvectors, labeled by the characters $\chi_k \in \hat{G}$, the character group of $G$, which satisfies $\hat{G} \cong G$. These eigenvectors, denoted $\ket{d_k}$, emerge as a direct consequence of the group grading structure of the category $\mathcal{C}_G$.

$\ket{d_k}$ is an eigenvector of $M_{\mathbf{g}}$ with eigenvalue $\lambda_k(\mathbf{g}) = \chi_k^{-1}(\mathbf{g}) \lambda(\mathbf{g})$. We conjecture these are the only possible eigenvectors of $M_{\mathbf{g}}$ with leading eigenvalues. Then, we have
\begin{align}
  \sum_{\hkh{q,s}} \prod_{m=1}^L d_{q_m}^{\frac12}
  = & \frac{1}{\cD^2} \sum_{k \in \hat{G}} \prod_{m=1}^L \chi_k^{-1}(\mathbf{g}_m) \lambda(\mathbf{g}_m)
  = \frac{1}{\cD^2} \sum_{k\in\hat{G}} \chi_k^{-1} \kh{\prod_{m=1}^L \mathbf{g}_m} \prod_{m=1}^L \lambda(\mathbf{g}_m) \\
  =&
  \begin{cases}
    \frac{1}{\cD^2_{\mathbf{1}}} \prod_{m=1}^L \lambda(\mathbf{g}_m) & \prod_{m=1}^L \mathbf{g}_m = \mathbf{1} \\
    0 & \prod_{m=1}^L \mathbf{g}_m \neq \mathbf{1}
  \end{cases}.
\end{align}
Notably, the condition $\prod_{m=1}^L \mathbf{g}_m = \mathbf{1}$ is required for a valid $G$-defect network $\hkh{\mathbf{g}}$.

Putting everything together, we obtain
\begin{equation}
  \mathcal{E}_A(\rho_{\hkh{\mathbf{g}}}) = -(L+1) \ln \cD_{\mathbf{1}}^2 + 2 \sum_{m=1}^L \ln \lambda(\mathbf{g}_m) + \cO(\euler^{-L}).
\end{equation}
After averaging over all possible $G$-defect networks, we have
\begin{equation}
  \mathcal{E}_A(\rho) = \kh{\frac{2}{\abs{G}} \sum_{\mathbf{g} \in G} \ln \lambda(\mathbf{g}) - \ln \cD_{\mathbf{1}}^2} L- \ln \cD_{\mathbf{1}}^2 + \cO(\euler^{-L}).
\end{equation}
And we find that TEN is still given by $\ten = \ln \cD_{\mathbf{1}}^2$.

\subsection{TMI}

To compute TMI, we evaluate the entanglement entropy of regions with different topologies, as prescribed by Eq.~\eqref{Eq_EM_CMI}.

For a region $A$, the reduced density matrix on $A$ is
\begin{align}
  \rho_A = \Tr_{\bar{A}} \rho = \frac{1}{\abs{G}^N} \sum_{\hkh{\mathbf{g}}} \Tr_{\bar{A}} \rho_{\hkh{\mathbf{g}}},
\end{align}
For a given $\hkh{\mathbf{g}}$, $\Tr_{\bar{A}} \rho_{\hkh{\mathbf{g}}}$ is still given by Eq.~\eqref{Eq_APP_SN_ReducedDensityMatrix}. However, a crucial difference from the TEN case is that we need to account for the number of $G$-defect network configurations.

We enumerate the number of distinct $G$-defect network configurations on a region $A$ with a fixed boundary condition. Let $\hkh{\mathbf{g}}_A$, $\hkh{\mathbf{g}}_{\partial A}$ and $\hkh{\mathbf{g}}_{\bar{A}}$ denote the collections of group elements on the edges in the interior of $A$, across the boundary $\partial A$, and in the complement $\bar{A}$, respectively. Let the region $A$ contain $N_A$ plaquettes, $E$ edges in the interior, $L$ edges across the boundary, $V$ vertices in the interior, and $\pi_0$ connected components. As we fix $\hkh{\mathbf{g}}_{\partial A}$, the total number of configurations on $A$ is initially $\abs{G}^E$. However, each vertex imposes a constraint, reducing the count by a factor of $\abs{G}^V$. Additionally, there are $\pi_0$ independent global relations among these constraints, effectively restoring $\abs{G}^{\pi_0}$ configurations. Therefore, the number of consistent $G$-defect networks on $A$ is
\begin{equation}
  \# = \abs{G}^{E-V+\pi_0}.
\end{equation}
Using the Euler characteristic $\chi_F$ of the region, given by $\chi_F = V - E + N_A$, we can rewrite the count as
\begin{equation}
  \# = \abs{G}^{N_A - \chi_F + \pi_0}.
\end{equation}
For a simply connected region, the Euler characteristic is $\chi_F=1$, yielding $\# = \abs{G}^{N_A}$. For an annular region, $\chi_F=0$, resulting in $\# = \abs{G}^{N_A+1}$.

Let $A$ be a simply connected region. For given $\hkh{\mathbf{g}}_A$ and $\hkh{\mathbf{g}}_{\partial A}$, the number of compatible $G$-defect network configurations $\hkh{\mathbf{g}}_{\bar{A}}$ in the annular region $\bar{A}$ is $\abs{G}^{N_{\bar{A}} + 1}$, where $N_{\bar{A}}$ is the number of plaquettes in $\bar{A}$. Thus, the reduced density matrix $\rho_A$ can be written as
\begin{align}
  \rho_A & = \frac{1}{\abs{G}^N} \sum_{\hkh{\mathbf{g}}} \Tr_{\bar{A}} \rho_{\hkh{\mathbf{g}}}
  = \frac{1}{\abs{G}^N} \sum_{\hkh{\mathbf{g}}_{\bar{A}}} \sum_{\hkh{\mathbf{g}}_A} \sum_{\hkh{\mathbf{g}}_{\partial A}} \frac{1}{\cD_{\mathbf{1}}^{2L-2}} \sum_{\hkh{q,s}} \kh{\prod_{m=1}^L d_{q_m}} \op{\hkh{\mathbf{g}}_{\partial A}, \hkh{q,s}} \\
  & = \frac{1}{\abs{G}^N} \abs{G}^{N_{\bar{A}}+1} \sum_{\hkh{\mathbf{g}}_A} \sum_{\hkh{\mathbf{g}}_{\partial A}} \frac{1}{\cD_{\mathbf{1}}^{2L-2}} \sum_{\hkh{q,s}} \kh{\prod_{m=1}^L d_{q_m}} \op{\hkh{\mathbf{g}}_{\partial A}, \hkh{q,s}}.
\end{align}
Here, we explicitly include the $G$-defect network $\hkh{\mathbf{g}}_{\partial A}$ to label the orthonormal states. The entanglement entropy for region $A$ can be evaluated as
\begin{align}
  S_A(\rho)
  = & -\abs{G}^{N_A} \sum_{\hkh{\mathbf{g}}_{\partial A}} \sum_{\hkh{q,s}} \frac{\prod_m d_{q_m}}{\abs{G}^{N_A + L - 1} \cD_{\mathbf{1}}^{2L-2}} \ln \frac{\prod_l d_{q_l}}{\abs{G}^{N_A + L - 1} \cD_{\mathbf{1}}^{2L-2}} \\
  = & -\sum_{\hkh{\mathbf{g}}_{\partial A}} \sum_{\hkh{q,s}} \frac{\prod_m d_{q_m}}{\abs{G}^{L-1} \cD_{\mathbf{1}}^{2L-2}} \ln \prod_l d_{q_l} + \sum_{\hkh{\mathbf{g}}_{\partial A}} \frac{1}{\abs{G}^{L-1}} \kh{\ln \abs{G}^{N_A+L-1} + \ln \cD_{\mathbf{1}}^{2L-2}} \\
  = & (\ln \abs{G}) N_A + \kh{\ln \abs{G} \cD_{\mathbf{1}}^2 - \frac{1}{\abs{G} \cD_{\mathbf{1}}^2} \sum_{\mathbf{g} \in G} N_{\mathbf{g}}} L- \ln \abs{G} \cD_{\mathbf{1}}^2.
\end{align}

Similarly, for $A$ being the disjoint union of two disks (like region $B$ in Fig.~\ref{Fig_TMIAnnulus}), we find
\begin{equation}
  S_A = (\ln\abs{G}) N_A + \kh{\ln \abs{G} \cD_{\mathbf{1}}^2 - \frac{1}{\abs{G} \cD_{\mathbf{1}}^2} \sum_g N_g} L - 2 \ln \abs{G} \cD_{\mathbf{1}}^2.
\end{equation}
For $A$ being an annulus (like region $ABC$ in Fig.~\ref{Fig_TMIAnnulus}), we find
\begin{equation}
  S_A = (\ln \abs{G}) N_A + \kh{\ln \abs{G} \cD_{\mathbf{1}}^2 - \frac{1}{\abs{G} \cD_{\mathbf{1}}^2}\sum_g N_g} L - \ln \abs{G} \cD_{\mathbf{1}}^4.
\end{equation}

Combining the results using Eq.~\eqref{Eq_EM_CMI}, we obtain
\begin{equation}
  \tmi = \frac12 (S_{AB} + S_{BC} - S_{B} - S_{ABC}) = \frac12 \ln \abs{G} \cD_{\mathbf{1}}^4
  = \ln \sqrt{\abs{G}} \cD_{\mathbf{1}}^2.
\end{equation}

\end{document}